\documentclass[ amssymb,amsmath,notitlepage]{revtex4-1}

\usepackage{dcolumn}
\usepackage{bm}
\usepackage[T1]{fontenc}
\usepackage{lmodern}

\usepackage{color}
\usepackage{graphicx}
\usepackage{epstopdf}
\usepackage{hyperref}

\begin{document}

\title{Compressive force generation by a bundle of living biofilaments}

\author{Sanoop Ramachandran}
\author{Jean-Paul Ryckaert}
\email{jryckaer@ulb.ac.be}
\affiliation{ 
 Physique des Polym\`eres, Universit\'e Libre de Bruxelles,\\
Campus Plaine, CP 223, B-1050 Brussels, Belgium
}%

\date{\today}

\begin{abstract}
To study the compressional forces exerted by a bundle of living stiff 
filaments pressing on a surface, akin to the case of an actin bundle in 
filopodia structures, we have performed particulate Molecular Dynamics 
simulations of a grafted bundle of parallel living (self-assembling) 
filaments, in chemical equilibrium with a solution of their constitutive 
monomers. 
Equilibrium is established as these filaments, grafted at one end to a wall 
of the simulation box, grow at their chemically active free end and encounter 
the opposite confining wall of the simulation box. 
Further growth of filaments requires bending and thus energy, which 
automatically limit the populations of longer filaments. 
The resulting filament sizes distribution and the force exerted by the bundle 
on the obstacle are analyzed for different grafting densities and different 
sub- or supercritical conditions, these properties being compared with the 
predictions of the corresponding ideal confined bundle model. 
In this analysis, non-ideal effects due to interactions between filaments and 
confinement effects are singled out. 
For all state points considered at the same temperature and at the same gap 
width between the two surfaces, the force per filament exerted on the opposite 
wall appears to be a function of a rescaled free monomer density 
$\hat{\rho}_1^{\rm eff}$.
This quantity can be estimated directly from the characteristic length of 
the exponential filament size distribution $P$ observed in the size domain 
where these grafted filaments are not in direct contact with the wall.
We also analyze the dynamics of the filament contour length fluctuations in 
terms of effective polymerization ($U$) and depolymerization ($W$) rates, 
where again it is possible to disentangle non-ideal and confinement effects. 
%
%
%
\end{abstract}

\maketitle
%
%
%
%
%
%
%
%
\section{Force generation by a bundle of parallel living filaments: 
Introduction\label{sec:intro}}
The generation of force and work by polymerizing actin filaments pushing on 
the cell membrane during the development of lamellopodes or filopodia 
or pushing on a host vesicle in the intracellular matrix (lysteria), has 
attracted much attention for the last 30 years.~\cite{b.Howard,Hill.81,Hill.82}
Biomimetic experiments have recently been set up to probe polymerizing 
filament networks on systems of controlled complexity which are designed 
to either probe the polymerization 
force directly~\cite{Dogterom.07, Brangbour.11} or to indirectly observe 
confinement effects on growing bundles.~\cite{Reymann.10}
The stationary force generated by a single actin filament polymerizing against 
a wall is established theoretically by invoking some form of ratchet 
mechanism~\cite{b.Howard,mogilner.96} for the 
monomer insertion between the tip of the filament and the wall. 
For a rigid filament hitting the wall normally, the standard result 
is~\cite{b.Howard}
\begin{equation}
f_N=\frac{k_{\rm B}T}{d} 
\ln{\left(\frac{\rho_1}{\rho_{\rm 1c}}\right)},
\label{eq:force}
\end{equation}
where $d$ is the incremental contour length per added monomer, 
$\rho_1$ the free monomer number density and $\rho_{\rm 1c}=1/K$ the free 
monomer critical density directly related to the effective equilibrium 
constant $K$ of the single monomer (de)polymerization reaction taking place 
at filament ends. 
This wall equilibrium force is the elementary case of the stalling force, 
namely the force needed to stop the progression of an arbitrary living 
filament network implying generally additional filament capping, severing 
or branching proteins and also the irreversible hydrolysis of ATP-actin 
complexes.

An interesting challenge of intermediate complexity is provided by the 
establishment of the force-velocity relationship for a bundle of parallel 
F-actin filaments which includes the determination of the force needed to 
stop the growth of the bundle in absence of any interference with additional 
coupled reactions involving other proteins. 

An experimental set up where the rigid polarized acrosome of Limulus sperm 
was used as an actin bundle initiator has been exploited~\cite{Dogterom.07} 
to measure the stationary force exerted by growing filaments on a fixed wall 
while simultaneously, the number of filaments implicated was independently 
determined.

The surprising result that the force for a eight-filament
bundle is essentially of the order of magnitude of the force in 
Eq.~(\ref{eq:force}),
expected theoretically for a single filament, has
been attributed to some dynamic instability-like length
fluctuations but the interpretation of this, 
unfortunately unique, experiment is still under 
debate.~\cite{Joanny.11,mogilner.09}
Theoretical rationalization in terms of stochastic models of filaments 
bundles~\cite{Joanny.11,Krawczyk.11,Frey.08} invariably suggest that, in 
supercritical conditions, the force exerted by a wall to stop a bundle of 
polymerizing filaments is extensive in the number of filaments. 
In these stochastic model approaches, some ad hoc choices must be made on 
how the total load force exerted through the wall on the bundle is shared by 
the individual filaments touching the wall. 
Similarly, there is no clear argument to decide how the wall force modifies, 
separately, the polymerizing rate $U$ and the depolymerizing rate $W$ when the 
tip of the filament is at contact. 
These microscopic aspects have however direct implications on the growth 
kinetics as it was shown explicitly for a dynamical model of a hundred 
filaments actin bundle pressing on a moving wall.~\cite{Joanny.11} 
The living filaments are treated as straight rods having fixed parallel 
orientation and are subject to discrete length fluctuations, some plausible 
rules being assumed for the (de)polymerizing rates relative to filaments 
hitting the wall, while the wall position itself is assigned at the tip of 
the longest filament(s) which can be unique or not, hence the need for the 
choice of an equal distribution of the total load force among them. 
Equation~(\ref{eq:force}) is recovered  for the stalling force per filament 
and the velocity-force dynamical behaviour shows a sharp transition between 
a non-condensed state at small load and a condensed state at higher loads 
where long filaments accumulate against the piston, the stalling regime being 
approached very slowly.~\cite{Joanny.11}
Another study including lateral attractive interactions between filaments 
suggests an increase, with respect to Eq.~(\ref{eq:force}), of the 
equilibrium force exerted by the bundle on the wall.~\cite{Krawczyk.11}
Next to these stochastic models which are either treated analytically (at the 
price of further approximations) or solved numerically using stochastically 
generated chemical events using Gillespie algorithm, attempts have been made 
to undertake a direct Brownian dynamics simulation of monomers at a mesoscopic 
level for the reacting system composed of filaments and a solution of individual
 free monomers. 
Along this approach where some ad-hoc rules are used to describe the 
self-assembly processes without an explicit consideration to fundamental 
(non-)equilibrium statistical mechanics aspects, one finds the case of a 
single living actin filament,~\cite{guo-09,guo-10} and the study of the 
self-assembly of living filament branched networks pressing on a mobile 
wall.~\cite{Lee.08,Lee.09}

Recently, we have proposed a new hybrid Molecular dynamics-Monte-Carlo 
simulation method sampling a reactive canonical ensemble for systems 
composed of a fixed number $N_f$ of semi-flexible filaments, modeled by a 
discrete wormlike chain model, and a solution of free monomers occupying a 
volume $V$ at temperature $T$.~\cite{Caby.12} 
Under the last ensemble constraint in the reactive system of a fixed total 
number of monomers $N_t$, single monomer (de)polymerizing events at filament 
ends are taking place along the microscopic time trajectory on the basis of 
an instantaneous Monte-Carlo spatially localized association/dissociation 
process satisfying micro-reversibility. 
With some proximity criterium and with some characteristic probability rate, 
a free monomer attempts an association step at an active end of a filament 
end while, in parallel, with some related probability rate, any filament end 
monomer attempts a depolymerization step. 
Any attempt, implying a topological change in the monomer connectivity within 
the reacting filament and a very limited spatial shift of the reacting monomer, 
is then subject to an acceptance probability based on the energy change implied.
 In case of acceptance, the time trajectory is then the object of a 
discontinuity related to the instantaneous modifications implied by the 
accepted chemical step. 
This method was successfully tested on a solution of living filaments with an 
artificial size upper limit to test the distribution of filament lengths at 
equilibrium for various thermodynamic conditions tuned by the set of 
independent state variables ($N_f,N_t,V,T$).

\begin{figure}
\includegraphics[scale=0.45]{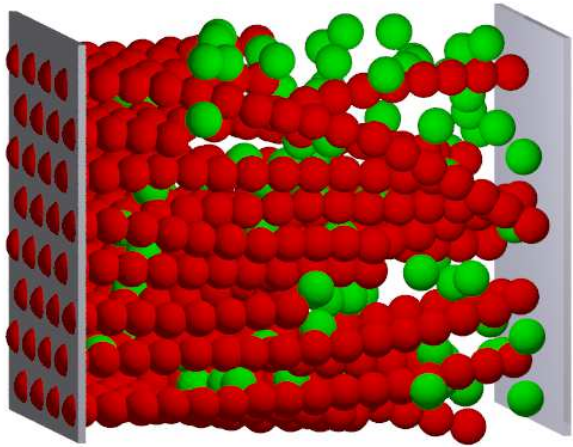}
\caption{Simulation box with $N_f$ living filaments (shown in red) anchored 
normal to the solid wall on the left, in chemical equilibrium with free 
monomers (shown in green). 
Single monomer end-filament (de)polymerization take place continuously for a 
prescribed total number $N_t$ of monomers. 
The filament growth (in supercritical conditions) is obstructed by the second 
solid wall on the right which exerts a normal equilibrium force on the bundle. 
The snapshot shown (note that periodic boundary conditions apply in lateral 
directions) is extracted from the $N_f=32,N_t=500$ case within the IIa 
experiments series (see Table~\ref{tab.1}).
\label{fig:setup}
}
\end{figure}

In the present paper, we exploit the above methodology to investigate the equilibrium state of a bundle of 
$N_f$ filaments of persistence length $l_{\rm p}$, grafted at one end to a flat
solid surface of area $A$ with their first segment constrained along the surface normal, as illustrated in Fig.~\ref{fig:setup}. 
A second wall, parallel to the first, with gap distance $L\ll l_{\rm p}$ is 
treated as an obstacle to the growth of the bundle's filaments by effective 
polymerization at their chemically active free end. Individual filaments have 
thus a contour length $L_{\rm c}$ limited beyond $L_{\rm c} \approx L$ 
by the larger and larger bending energy penalty needed to accomodate longer 
and longer filament sizes. 
The Hamiltonian model used is similar to the one exploited for free living 
filaments in solution~\cite{Caby.12} with the same intramolecular model and 
the same treatment of the intermolecular forces but, in the present case, it 
includes a confinement potential and grafting constraints. 
The whole system, treated in the canonical reactive ensemble specified by the 
independent variables ($N_f,N_t,A,L,T$), furnishes a variety of thermodynamic, 
structural and dynamical properties which are detailed in the bulk of this 
paper.

In the ideal solution approximation where the intermolecular forces in the 
Hamiltonian are simply turned off, the structural and thermodynamical 
properties of the same grafted and confined bundle reactive system considered 
within the same ($N_f,N_t,A,L,T$) canonical reactive ensemble can be obtained 
analytically, except for single filament-wall integrals which generally do 
require single filament Monte-Carlo calculations.~\cite{RRMP.13} 
This illuminating ideal solution treatment of such a complex system has been 
developed in detail separately,~\cite{RRMP.13} but is however the object of a 
short summary in the present paper, given the important role it plays in the 
guidance of the massive particulate simulations reported here for the first time
 and in the comparative analysis with the results of corresponding simulations 
of the non-ideal case. 
Let us just anticipate here that this theoretical treatment furnishes, in the 
ideal solution apprximation, the filament size distribution and the related 
average force exerted by such a living filament on the obstacle wall. 
These properties strongly rely on the already mentionned filament-wall 
integrals which deal with the effective wall force and associated mean force 
potential between a fixed size grafted filament and an obstacle wall located 
at distance L from the filament grafted end and oriented perpendicularly to 
the initial end-grafting direction. 
On the specific single grafted filament-wall topics, let us mention an 
illuminating study of the interaction between a grafted continuous wormlike 
chain and a hard wall where universal relations, valid in the stiff 
limit where $l_{\rm p}\gg L$, are reported.~\cite{Frey.06}

The paper is organized as follows: in the first part of Sec.~\ref{sec:thermo}, 
we describe a general thermodynamics and statistical mechanics framework in 
which the properties of our grafted and confined bundle of interacting 
living filaments can be formulated. 
We introduce various quantities which are needed to quantify non-ideal effects 
and confinement properties and define the filament osmotic force concept. 
In the second part of the same section, we report the 
most important general properties of the ideal bundle system. 
The next section, Sec.~\ref{sec:conf-bundle}, is concerned with the explicit 
microscopic model which is used in our particulate simulations. 
We first concentrate in Sec.~\ref{sec:hamil1} on the single fixed size 
filament model, including its interaction with the obstacle wall while the 
global system aspects of the model, including the living filament aspects, 
are reported in Sec.~\ref{sec:hamil}. 
We finally give in Sec.~\ref{sec:simul_exp} the list of adopted 
parameters.
In Sec.~\ref{results}, we start by giving the list of experiments organized by 
increasing grafting density and increasing total number of monomers for a 
unique set of the other independent variables $(L,T)$. 
Results are then presented, regrouped into specific subsections devoted to 
the different structural and dynamical investigated properties. 
Finally, Sec.~\ref{sec:disc} contains an overall discussion and indicates 
some perspectives.

\section{Thermodynamic description of the confined bundle\label{sec:thermo}}

\subsection{The general approach \label{sec:thermo1}}

The bundle model detailed in previous section can be described 
thermodynamically in terms of the Helmholtz free energy $F(T,A,L,N_t,N_f)$ of a 
closed reacting fluid system at temperature $T$, having a fixed total number
$N_t$ of solute monomers enclosed in a slit pore of transverse area $A$ and gap 
width $L$ and having $N_f$ permanent filament seeds anchored at one wall giving 
rise to a bundle of living filaments. 
These filaments undergo single monomer exchanges at their free end with the
free monomers in the bath according to reversible chemical reactions
\begin{equation}
A_{i-1} + A_1 \rightleftharpoons A_{i}      \ \ \ (3<i \leq z^{*})
\label{eq:reaction}
\end{equation}
where $A_i$ and $A_1$ represent a grafted filament of size $i$ 
and a free monomer, respectively . The length of a filament of size $i$ is 
$L_{{\rm c}i}=(i-1)d$ where $d$ is the monomer size, more exactly the filament 
contour length contribution per inserted monomer.
This series of reactions (and series of chemical species) is considered as 
limited to the indicated size window in our theoretical framework. 
The choice of a minimum size of three monomers is motivated by the fact that 
the chemical equilibrium constant associated to the reaction 
Eq.~(\ref{eq:reaction}) becomes size independent typically beyond 
$i=3$.~\cite{b.Howard} 
The neglect of shorter assemblies will be shown to have no impact on our 
conclusions, a fortiori in supercritical conditions where short filament 
populations are marginal.
The upper limit in the distribution, defined as
\begin{equation}
z^{*}=\frac{\pi L}{2 d},
\label{eq:z*}
\end{equation}
prevents the occurrence of grafted filaments which, after an initial large 
amplitude bending, would start growing without further bending energy, 
parallel to the the obstacle wall. 
In Eq.~(\ref{eq:z*}), $z^{*}$ corresponds to the size of a filament which can 
be mapped on a quarter of a circle of radius $L$, where $L$ is the gap width 
between the grafting and obstacle planes. 
Focusing in the present work to the compressive force exerted by a bundle of 
relatively stiff filaments ($l_{\rm p} \gg L$), it will be verified later that 
filaments of size $z^{*}$ remain practically unpopulated in the supercritical 
regime as long as the free monomer concentration is appropriately limited. 
A quantitative assessment of this limit will be given in Eq.~(\ref{eq:rho1lim}).

Starting with the general differential form of the free energy for a non-
reactive mixture of all species with $N_1$ free monomers, $N_3$ trimer filaments, $N_4$ quadrimer filaments $\dots$, treated as 
independent variables, one has 
\begin{align}
dF= -SdT-p_N A dL&-p_T L dA+\mu_1 dN_1+\sum_{i=3}^{z^{*}} \mu_{i}dN_i.
\label{eq:dF1}
\end{align}
Here $p_N$ and $p_T$ represent the normal and the tangential 
pressures respectively and the $\mu$'s are the imposed chemical potentials of
 the various species.
If we now relax the constraint of non-reactive mixture and impose the 
chemical equilibrium for all reactions in Eq.~(\ref{eq:reaction}), one must 
have
\begin{equation}
\mu_{i-1} + \mu_1 = \mu_{i} \ \ \ (3 <i \leq z^{*}).
\label{eq:mu}
\end{equation}
At the same time, we require that the relevant composition variables,
$N_t$ and $N_f$, are related to species composition variables by
\begin{align}
N_t&= N_1+3 N_3+ 4 N_4+ 5 N_5+\ldots+ z^{*}N_{z^{*}}
\label{eq:constr1}
\end{align}
and
\begin{align}
N_f&= N_3+ N_4+ N_5 + \ldots +N_{z^{*}}.
\label{eq:constr2}
\end{align}
Taking into account Eqs.~(\ref{eq:mu}),~(\ref{eq:constr1}) 
and~(\ref{eq:constr2}), Eq.~(\ref{eq:dF1}) becomes
\begin{align}
dF= -SdT-&p_N A dL-p_T L dA+\mu_1 dN_t \nonumber\\
&+(\mu_3-3 \mu_1)dN_f .
\label{eq:dF2}
\end{align}

In terms of the adopted independent variables, the normal pressure $p_N$ 
and the free monomer chemical potential $\mu_1$ are thus formally given by
\begin{align}
p_N&= -\frac{1}{A} \left(\frac{\partial F}{\partial L}\right)_{N_t,N_f,A,T},
\label{eq:pN}\\
\mu_1&= \left(\frac{\partial F}{\partial N_t}\right)_{N_f,A,L,T}.
\label{eq:mu1}
\end{align}
By analogy with in vitro experimental conditions where the confined bundle 
in solution is in thermodynamic and chemical equilibrium with an effective 
large reservoir of a free monomer solution at the same temperature $T$ and same 
monomer chemical potential $\mu_1$, it is useful to identify the pressure 
exerted by the filament brush on the opposite wall, not by the total pressure 
$p_N$ defined in Eq.~(\ref{eq:pN}), but by the excess (osmotic) pressure 
$\Pi$ defined by
\begin{equation}
\Pi=p_N-p^{\infty}(\mu_1,T)
\label{eq:pi}
\end{equation}
where $p^{\infty}(\mu_1,T)$ is the isotropic pressure in the reservoir fluid 
(free monomers) with specific number density $\rho_1^{\infty}(\mu_1,T)$.
The activity coefficient $f_1^{\infty}$ of this reservoir solution which 
takes into account all interactions between the free monomers is defined by 
the expression using the ideal solution as reference state,
\begin{align}
\beta \mu_1&=
-\ln{(q_1/V)}+\ln{(f_1^{\infty} \rho_1^{\infty})}, \nonumber \\
&=\ln{\Lambda^3}+\ln{(f_1^{\infty} \rho_1^{\infty})}
\label{eq:mu1*}
\end{align}
where $q_1$ is the ideal gas single monomer partition function at same 
temperature $T$ and $\Lambda$ is the de Broglie wave length associated to the 
free monomers.

Considering all permanent grafting points as equivalent, we can define the 
activity coefficient $f_i$ of a grafted filament of size $i$ in the 
brush/solution system at temperature $T$ as 
\begin{equation}
\beta \mu_i=-\ln{q_i}+\ln{(f_i P_i)}=-\ln{(q_i^0 \alpha_i)}+\ln{(f_i P_i)}
\label{eq:mufili}
\end{equation}
where $q_i(T,L)$ is the ideal solution partition function at temperature $T$ 
of a single chemisorbed filament of size $i$ in interaction (if sufficiently 
long) with the opposite obstacle wall and where $P_i=N_i/N_f$ is the number 
fraction of filaments of size $i$. 
In the second equality of Eq.~(\ref{eq:mufili}) we introduce the ideal solution 
partition function $q_i^0(T)$ for the same grafted filament of size $i$ in 
absence of the obstacle wall. 
The wall factor $\alpha_i(L,T)$ is given by
\begin{equation}
\alpha_i=\frac{q_i}{q_i^0}=\langle \exp{(-\beta U^{\rm w})}\rangle_i^0
\label{eq:wallfac}
\end{equation}
where $\langle \ldots \rangle_i^0$ is a canonical average 
($\beta=(k_{\rm B}T)^{-1}$) based on a single grafted filament Hamiltonian 
$H_i^0$ for a filament of size $i$ in absence of wall interaction term 
$U^{\rm w}$. 
These wall factors are trivially equal to unity as long as the single grafted 
filament does not hit the opposite wall, which defines an upper limit $z$ for 
the series of filaments avoiding any direct interaction with the wall which 
should be of the order $z \approx L/d$.

Combining Eqs.~(\ref{eq:mu}),~(\ref{eq:mu1*}) and~(\ref{eq:mufili}), 
one gets~\cite{b.Hill}
\begin{equation}
K_{0i}\equiv \frac{q_i}{q_{i-1}\Lambda^{-3}}
=\frac{f_i}{f_{i-1} f_1^{\infty}} \frac{P_i}{P_{i-1}\rho_1^{\infty}}, 
\label{eq:equi_const1}
\end{equation}
where we introduce the equilibrium constant $K_{0i}$ of the reaction in
Eq.~(\ref{eq:reaction}) within the reference ideal system, a quantity 
depending only on the temperature (and possibly $L$). In terms of wall factors, 
the ideal solution equilibrium constant can be reformulated as 
\begin{equation}
K_{0i}= \frac{q_i^0}{q_{i-1}^0\Lambda^{-3}}\frac{\alpha_i}{\alpha_{i-1}}
= K_0 \frac{\alpha_i}{\alpha_{i-1}}
\label{eq:equi_const2}
\end{equation}
where we have assumed~\cite{b.Howard} that the (de)polymerizing equilibrium 
constant of the active free end of the filament in bulk, noted $K_0$, 
is independent of the filament size. 

Finally, the special form of the chemical potential Eq.~(\ref{eq:mufili}) 
which disentangles wall and interaction effects, lead to filament 
populations satisfying a set of equalities
\begin{equation}
\frac{P_i}{P_{i-1}}=\rho_1^{\infty} K_{0} \frac{\alpha_i}{\alpha_{i-1}} 
\frac{f_{i-1} f_1^{\infty}}{f_i}  \ \ \ (3<i \leq z^{*})
\label{eq:equi_const3}
\end{equation}
For the ideal bundle case, these relations take the form
\begin{equation}
\frac{P_i^{\rm id}}{P_{i-1}^{\rm id}}=\rho_1 K_{0} \frac{\alpha_i}{\alpha_{i-1}} \ \ \ (3<i \leq z^{*})
\label{eq:equi_const4}
\end{equation}
where $\rho_1$ is the free monomer density.

\subsection{The ideal bundle properties
\label{sec:thermo2}
}

The statistical mechanics properties of the ideal bundle case in 
the $(N_t,N_f,A,L,T)$ canonical reactive ensemble was treated 
in detail elsewhere.~\cite{RRMP.13}
In terms of the reduced equilibrium free monomer density 
$\hat{\rho}_1=\rho_1 K_0$, the filament number fractions in the ideal solution 
at equilibrium are given by
\begin{align}
P_{i}^{\rm id}&=\frac{N_i}{N_f} 
= \frac{(\hat{\rho}_1)^{i}}{D} 
\; \; \; 
(3 \leq i \leq z), \label{eq:popi}\\
P_{z+k}^{\rm id}&= \frac{N_{z+k}}{N_f} 
= \alpha_{z+k}(L) \frac{(\hat{\rho}_1)^{z+k}}{D} 
\; \; \; (1 \leq k \leq k^{*}),
\label{eq:pop}
\end{align}
where the normalization factor is given by
\begin{equation}
D = \left[\sum_{i=3}^z 
(\hat{\rho}_1)^{i}\right] 
+ \sum_{k=1}^{k^{*}} \alpha_{z+k}(L) (\hat{\rho}_1)^{z+k}. 
\label{eq:ddd}
\end{equation}
Equation~(\ref{eq:popi}), valid for filaments too short to interact with the 
wall, is an exponential size distribution 
$\propto \exp{(i/s)}$ with $s=\left[\ln{\hat{\rho}_1}\right]^{-1}$, 
where $\left|s\right|$ is the characteristic length scale (in monomer units) 
of the growing ($\hat{\rho}_1>1$) or decreasing ($\hat{\rho}_1<1$) exponential 
distribution. 
Equation~(\ref{eq:pop}) deals with filaments of size $z+k$ hitting the wall, 
where $k$ is limited by $k^{*}=z^{*}-z$. 
Equations~(\ref{eq:popi}) and~(\ref{eq:pop}) are compatible with 
Eq.~(\ref{eq:equi_const4}). 
To express this filament distribution in terms of the original independent 
variables of the reactive canonical ensemble, one substitutes the 
filament densities Eqs.~(\ref{eq:popi}) and~(\ref{eq:pop}) in the constraint 
relationship Eq.~(\ref{eq:constr1}) and gets the $\rho_1$ implicit equation
\begin{equation}
\rho_t= \rho_1+\frac{\sigma_f}{L} \langle i \rangle
=\rho_1+\frac{\sigma_f}{L} \frac{M(\hat{\rho}_1,L)}{D(\hat{\rho}_1,L)},
\label{eq:rho1}
\end{equation}
where $\rho_t$ is the total monomer number density, $\sigma_f$ is the 
filament surface density and $\langle i \rangle$ is the average 
length of the filaments in the bundle. 
In Eq.~(\ref{eq:rho1}), $D$ is given by Eq.~(\ref{eq:ddd}) and 
$M=\hat{\rho}_1 \partial D/\partial \hat{\rho}_1$.

In Ref.,~\cite{RRMP.13} by applying Eq.~(\ref{eq:pN}) to the theoretical 
expression of the free energy of the ideal bundle solution, we get the normal
pressure $p_N$ on the obstacle wall as
\begin{align}
\beta p_N =  \rho_1 
+ \sigma_f \left(\frac{\partial \ln{D}}{\partial L}\right)_{\hat{\rho}_1}.
\label{eq:paint}
\end{align}
Using Eq.~(\ref{eq:pi}) with $p^{*}=\rho_1 k_{\rm B}T$, one gets 
the force per filament $f_N^{\rm id}$ in an ideal solution as
\begin{align}
\beta f_N^{\rm id}&= 
\frac{\Pi}{\sigma_f k_{\rm B}T} 
= \left(\frac{\partial \ln{D}}{\partial L}\right)_{\hat{\rho}_1}
=\sum_{k=1}^{k^{*}} \beta \bar{f}_{z+k} P_{z+k}^{\rm id}.
\label{eq:fn}
\end{align}
In Eq.~(\ref{eq:fn}), the new force $\bar{f}_{z+k}$ appearing in the last 
term derives from a potential of mean force 
\begin{align}
w_{z+k}(L)&\equiv -k_{\rm B}T \ln{(\alpha_{z+k}(L))},
\label{eq:potmf}
\end{align}
and thus represents the force exerted by the wall on a filament of fixed contour
 length $z+k$, averaged over all the microscopic degrees of freedom associated 
to its single grafted filament Hamiltonian $H_{z+k}$.
According to Eq.~(\ref{eq:fn}), the equilibrium normal force $f_N^{\rm id}$ is 
the weighted average of the force $\bar{f}_{z+k}$, using absolute probabilities 
given by Eq.~(\ref{eq:pop}), over all hitting filament species denoted by 
$k=1,k^{*}$.
The chemical conditions are fixed by $\hat{\rho}_1$, while all structural 
aspects including the persistence length and the precise nature of the 
repulsive wall potential $U^{\rm ext}$ are absorbed in the $L$ dependence of 
the different $\alpha_{z+k}(L)$ wall factors given by Eq.~(\ref{eq:wallfac}). 
This is equally true for $p_N$ in Eq.~(\ref{eq:paint}), which represents the 
force per unit area originating from independent living filaments defined by 
the same specifications.
These wall factors, which are responsible for the size distribution of grafted 
living filaments hitting a wall perpendicular to their initial grafting 
orientation, have been shown~\cite{Frey.06} to present universal relations for 
the model of a continuous wormlike chain (WLC) model hitting a hard wall, as 
long as $L/l_{\rm p}<0.1$. 
The explicit universal relations express 
$\alpha \equiv \tilde{Z}(\tilde{\eta})$ in terms of the rescaled compression 
variable
\begin{align}
\tilde{\eta}=\frac{L_{\rm c}-L}{L_{\parallel}}=\frac{(L_{\rm c}-L) l_{\rm p}}{L_{\rm c}^2}
\label{eq:etatilde}
\end{align}
where $L_{\rm c}$ is the contour length of the filament and $L_{\parallel}$ a 
characteristic length. 
For other filament/wall models or for more flexible WLC, the derivation of the 
wall factors require Monte-Carlo calculations, as discussed in 
references.~\cite{Frey.06,RRMP.13}

The theoretical developments of this work require that filaments are 
sufficiently rigid over the gap size $L$ to prevent filaments to grow laterally.
 This was formulated by requiring that the thermodynamic conditions guarantee 
that $P_{z^{*}}\cong 0$, where $z^{*}$ is given by Eq.~(\ref{eq:z*}). 
Guided by the ideal bundle results and the universal properties of the wall 
factor,~\cite{Frey.06} it can be shown~\cite{RRMP.13} that if $L/l_{\rm p}<0.1$, 
the reduced density must remain below a limit value $\rho_{1b}$, namely
\begin{align}
\ln{\hat{\rho}_1} < \frac{l_{\rm p} d}{L^2}\equiv 
\ln{\hat{\rho}_{1b}}.
\label{eq:rho1lim}
\end{align}
This point was also carefully discussed in Ref.~\cite{Dogterom.07} dealing 
with the measurement of the compressive force of a bundle of stiff actin 
filaments. 
The data exploited in that study were obtained in conditions where 
Eq.~(\ref{eq:rho1lim}) applies.

\section{The specific model with intermolecular interactions for the confined 
self-assembling grafted bundle
\label{sec:conf-bundle}}

\subsection{Model Hamiltonian for a grafted and confined single filament and 
single living filament properties \label{sec:hamil1}}

Let us first describe the basic ingredients of the model, namely the way a 
filament of size $N$ is built as a linear assembly of $N$ point particles 
of mass $m$ with Cartesian coordinates $\left\{{\bf r}_i\right\}$ and 
momenta $\left\{{\bf p}_i\right\}$ with $i=1, \dots, N$. To graft the filament 
to the $x=0$ plane, the two first monomers are fixed in space by two independent
 infinitely stiff springs to positions 
(${\bf r}_1\equiv(0,y_{\gamma}, z_{\gamma})$, 
${\bf r}_2\equiv(d,y_{\gamma}, z_{\gamma})$) defining the first rigid bond of 
length $d$ of the filament at location $\gamma$ of the grafting surface, with 
normal orientation with respect the latter.
The adopted single filament Hamiltonian is
\begin{align}
H_N({\bf r},{\bf p})&= 
\sum_{i=1}^N \frac{{\bf p}_i^2}{2 m} -(N-1) \epsilon_0' 
+ \sum_{i=2}^{N-1} \frac{k_s}{2}(d_i-d)^2
\nonumber\\
&+ \frac{\kappa}{d} \sum_{i=2}^{N-1}(1-\cos \theta_i)
+\frac{N-2}{2 \beta} \ln{\frac{2 \pi}{\beta k_s d^2}}
\nonumber\\
&+U_N^{\rm ext},
\label{eq:hamil}
\end{align}
which starts with the kinetic energy term. 
The second term expresses the bonding energy corresponding to the energy 
released as heat when a new monomer attaches the filament and forms a new 
bond. 
There are $N-2$ bonds ${\bf d}_i={\bf r}_{i+1}-{\bf r}_{i}$ of (almost) 
constant length $d$ (the first bond length is automatically fixed by the 
grafting conditions) as we consider in the third term stiff harmonic springs 
with $d_i$ supposed to oscillate harmonically around $d$. 
The next term accounts for the bending energy where $\theta_i$ is the 
bending angle between bonds ${\bf d}_{i-1}$ and ${d}_{i}$, while the bending 
modulus $\kappa$ fixes the persistence length $l_{\rm p}$ of the filament 
according to $\kappa=k_{\rm B}T l_{\rm p}$. 
Our interest lies in situations where $l_{\rm p} \gg L$. 

The constant fifth term is useful to normalize the 
$\exp\left(-\beta k_s(d_i-d)^2/2\right)$ term which will appear in the 
filament canonical partition function (or in any phase space integral) 
in order to allow the (scalar) Gaussian bond length distribution around 
the mean $d$ to properly evolve towards the delta function in the 
$k_s \rightarrow \infty$ limit (harmonic bond stiff limit). 
In Ref.,~\cite{Caby.12} the Hamiltonian was defined without the fifth 
normalizing term now appearing in Eq.~(\ref{eq:hamil}). 
For consistency, the bonding energy $\epsilon_0$ in Ref.~\cite{Caby.12} 
must then be linked to the present bonding energy parameter using~\cite{RRMP.13}
\begin{equation} 
\beta \epsilon_0'
=\beta \epsilon_0+\frac{1}{2} 
\ln{\left(\frac{2 \pi}{\beta k_s d^2}\right)}.
\end{equation}
The external potential term which controls the interaction of the filament with 
the obstacle wall is represented as 
$U_N^{\rm ext}=\sum_{i=3}^N U^{\rm w}(L-x_i)$, where among all monomer-wall 
interactions, only terms with distance to the wall within the cut-off distance 
$x_{\rm c}$ ($L-x_i<x_{\rm c}$) contribute. 
We adopt the specific form
\begin{align}
U^{\rm w}(s)=\frac{3 \sqrt{3}}{2} 
\epsilon_{\rm w} 
\left[
\left(\frac{\sigma_{\rm w}}{s}\right)^9
-\left(\frac{\sigma_{\rm w}}{s}\right)^3 \right]
+ \epsilon_{\rm w},
\label{eq:walpot}
\end{align}
where $s=L-x$ is the distance between a monomer center and the wall. 
The cutoff distance is chosen at the minimum of the 
potential, hence $x_{\rm c}=3^{1/6} \sigma_{\rm w}$. 

In Eq.~(\ref{eq:mufili}), we introduced two forms of the canonical partition 
function at temperature $T$ of a single filament of size 
$3 \leq i \leq z^{*}$ grafted to one wall, namely $q_i(L,T)$ and $q_i^0(T)$ to 
be distinguished respectively by the inclusion or not of confining opposite 
wall $U^{\rm w}$ terms. 
These partition functions, now fully determined by 
the single filament Hamiltonian Eq.~(\ref{eq:hamil}) including the wall 
potential, Eq.~(\ref{eq:walpot}), can be exploited to compute two relevant 
quantities needed to establish the link between our non-ideal simulations with 
the ideal bundle case. 
Using Eq.~(\ref{eq:wallfac}), the wall factors $\alpha_i(L,T)$ can be 
determined by Monte-Carlo sampling \cite{RRMP.13}. 
Next, according to Eq.~(\ref{eq:equi_const2}), the equilibrium constant 
$K_0$ of the (de)polymerization reaction Eq.~(\ref{eq:reaction}) in absence of 
wall interference can be established and one gets, as expected, an 
$i$-independent equilibrium constant\cite{RRMP.13}
\begin{align}
K_{0}&=\frac{q_i^0}{q_{i-1}^0}\Lambda^3 = \frac{ 2\pi d^4}{l_{\rm p}}\;
\exp{(\beta \epsilon_0')}\; 
\left[1-\exp{(-2l_{\rm p}/d)}\right] F(w_0)
\label{eq:K0}
\end{align}
where the last term is a correcting factor for the bond flexibility function 
of $w_0=d/\sigma_d=\sqrt{\beta k_s d^2}$ which is explicitly given by 
\begin{align}
F(w_0)=
\frac{
\frac{1+{\rm erf}[w_0]}{2}(1+w_0^2)+ 
\frac{w_0}{\sqrt{2 \pi}} \exp{(-\frac{w_0^2}{2})}
}{w_0^2},
\label{eq:fw0}
\end{align} 
where the error function erf is used. In the stiff bond limit(discrete WLC 
model), one trivially observes that $\lim_{k_s \rightarrow \infty}F(w_0)=1$. 

\subsection{Model Hamiltonian for a bundle of self-assembled filaments and 
free monomers in interaction and the corresponding reactive canonical 
ensemble\label{sec:hamil}}

We will specifically perform Molecular Dynamics simulations with explicit 
chemical steps in order to probe the reactive canonical partition function 
$Q^{\rm RC}(N_t,N_f,A,L,T)$ relative to a bundle of $N_f$ grafted filaments 
of variable length confined in a slit pore of gap width $L$ and  transverse 
square area $A=H^2$, in thermal contact with a thermostat at temperature $T$. 
The $N_f$ filaments are grafted at the vertices, denoted by index 
$\gamma=(1,N_f)$, of a centered square lattice with unit cell length $a$ 
adjusted to get a predefined surface density $\sigma_f=N_f/H^2=2/a^2$. 
The pore encloses in total $N_t$ monomers which can either be independent 
free monomers or integrated within a grafted self-assembled filament. 

The partition function $Q^{\rm RC}$ is a sum of standard canonical partition 
functions over all distinguishable ways to group the $N_t$ monomers into 
topologies satisfying the global constraints 
Eqs.~(\ref{eq:constr1}) and~(\ref{eq:constr2}), where to each chemisorption 
site $\gamma$ is associated a specific filament size $i(\gamma)$ in the allowed 
range ($3 \leq i(\gamma) \leq z^{*}$) while all remaining monomers are treated 
as free monomers (the explicit form of $Q^{\rm RC}$ is mentioned in Appendix A). 
To each topology is associated an Hamiltonian $H^{\rm tot}$ and corresponding 
canonical ensemble partition function. This Hamiltonian can be formally written 
as the sum of all single filament Hamiltonians $H_N$ (Eq.~(\ref{eq:hamil}) 
with $N=i(\gamma)$ for any site $\gamma$) and all single free monomer 
Hamiltonian $h_k$
\begin{align}
h_k&= \left[\frac{\textbf{p}_k^2}{2 m}+U^{\rm w}(x_k)
+U^{\rm w}(L-x_k)\right]
\end{align}
grouping kinetic and wall interaction potentials, to which intermolecular 
interactions must be added. One thus gets
\begin{align}
H^{\rm tot}&=\sum_{\gamma=1}^{N_f}H_{i(\gamma)}+ \sum_{k=1}^{N_1} h_k 
+ U^{\rm ev}
\label{eq:hamtot}
\end{align}
where $N_1=N_t-\sum_{\gamma=1}^{N_f} i(\gamma)$ and where the excluded volume 
interactions term $U^{\rm ev}$ is taken as in our previous 
study,~\cite{Caby.12} as the sum of purely repulsive interactions 
(Weeks-Chandler-Andersen (WCA) potential) between all pairs of monomers 
except between pairs of monomers belonging to the same filament. 

The exploration of the full reactive ensemble is performed during the Molecular 
Dynamics trajectory by performing in addition (according to stochastic rules) 
topological changes through explicit polymerization steps (when a free monomer 
is close to a filament free end) or through depolymerization steps. 
The rules were defined in Ref.~\cite{Caby.12} and are trivially adjusted for 
the present study by adding the wall potential terms to the 
excluded volume potential terms for the reacting monomer in the Monte-Carlo 
chemical step procedure. 
Consequently, a depolymerization step where the freed monomer is relocated at a 
place of to high wall potential energy, will typically be refused. 
Also, for the polymerization step, the availability of free monomers in the 
reactive volume (close to the tip of the receiving filament) to attempt a 
polymerization step will be lowered by the wall presence when the filament tip 
precisely hits the wall.

\subsection{Computer simulation experiments: Choice of parameters and list of 
experiments\label{sec:simul_exp}}

A bundle of $N_f=32$ filaments is grafted at one wall of square area $A=H^2$, 
the filaments seeds being permanently linked to the sites of a centered square 
lattice of unit cell size $H/4$. 
A free monomer solution fills the gap of width $L$ between the grafting plane 
and the obstacle parallel plane while periodic boundary conditions are used in 
transverse directions, giving rise to a slit pore confining volume $V=AL$. 
The hybrid Molecular Dynamics - Multiparticle Collision Dynamics - Monte-Carlo 
(MD-MPCD-MC) method follows dynamically the system of $N_t$ monomers according 
to the Hamiltonian, Eq.~(\ref{eq:hamtot}), and performs in addition local 
(de)polymerization steps, thus sampling the reactive canonical ensemble 
associated to our system at fixed ($N_t,N_f,A,L,T$). 
The solute system of $N_t$ monomers is coupled to a MPCD bath, the temperature 
being explicitly fixed by its value in the acceptance rules of the 
Monte-Carlo chemical steps.~\cite{Caby.12} 
The side $a_0$ of the elementary MPCD (local) collisional cubic boxes is taken 
as twice the monomer size increment $d$ of the filaments.~\cite{Caby.12} 
The simulation box has its $L$ and $H$ dimensions corresponding to an integer 
number of $a_0$, all experiments having been performed for the same gap width 
$L=8a_0=16d$. 
Table~\ref{tab.1} summarizes the specific parameters of the different series 
of experiments denoted by roman numbers I-III ordered by increasing filament 
brush surface densities but differing also by the choice of the ideal solution 
equilibrium constant $K_0$. Experiments I and IIb adopt the same $K_0$ value, 
while this constant is halved for experiment IIa and again halved for experiment
 III. 
Each of the four series of experiments are performed for several values for 
the total number of monomers, $N_t$ to cover various sub- and supercritical 
regimes. 

We now list the parameters adopted in our simulations using $k_{\rm B}T$ 
as energy unit, the MPCD solvent mass $m_{\rm s}$ as unit of mass, and finally 
$a_0$ as unit of length. The time unit is $u_t=a_0 \sqrt{m_{\rm s}/k_{\rm B}T}$.
The monomers (either free or part of the filaments) have a mass 
$m=5m_{\rm s}$. The bond length at potential minimum $d=0.5 a_0$ will also be 
regularly used when mentioning results in reduced variables implying length 
scales. 
The bonding force constant is taken to be $k_s=400 k_{\rm B}T/d^2$. 
The bending potential is fixed by $l_{\rm p}/d=250$, with 
$\kappa=k_{\rm B} T l_{\rm p}/d$. 
The bonding energy parameter $\epsilon_0'$ varies for the different sets of 
experiments (see Table~\ref{tab.1}). For experiment IIa, we take  $\epsilon_0'=7.34894$, which gives 
$K_0=4.88373 a_0^3$ according to Eq.~(\ref{eq:K0}) 
(note that this set of parameters is fully equivalent to the one used in the 
previous work where we took $\epsilon_0=9.42573$ in Eq.~(31) 
in Ref.~\cite{Caby.12}.)
The Lennard-Jones (LJ) parameters for the WCA pair interaction is set 
to $\sigma=0.44545 a_0$ and $\epsilon=3 k_{\rm B}T$, while for the wall 
potential $U^{\rm w}$ in Eq.~(\ref{eq:walpot})
the two parameters were set to 
$\sigma_{\rm w}=0.5 a_0=d$ and $\epsilon_{\rm w}=0.1 k_{\rm }T$ and the cutoff distance (chosen at minimum of potential) is 
$x_{\rm c}=3^{1/6} \sigma_{\rm w} =1.200936 d$. 
The MPCD parameters are the solvent density $\rho_{\rm s}=5 a_0^{-3}$, 
the collision time interval $\Delta t=0.1 u_t$, and the collision rotation 
angle of $\alpha=130$ degrees.  
The equations of motion for solute filaments and free monomers are integrated 
using the velocity version of the Verlet algorithm with MD time step of 
$h=0.002 u_t$. 
The last parameter to be fixed is the attempt frequency $\nu$ with which any 
reactive monomer (polymerizing or depolymerizing) is subject to an attempted 
chemical step, the central parameter controlling the time scale on which the 
filament contour length dynamics takes place.~\cite{Caby.12} 
With respect to our previous work, we took $\nu=50 u_t^{-1}$ instead of 
$5$ to accelerate the establishment of chemical equilibrium. 

\section{Simulation results \label{results}}

Simulation data were obtained, for each state point investigated, by 
averaging over four independent MD-MPCD-MC runs of $T_{\rm run}=10^5 u_t$ which 
represents one order of magnitude longer than the single filament contour 
length relaxation time $\tau_i$ (see Table   \ref{tab.2}).

Each run was preceded by an equilibration run of $T_{\rm run}=10^4 u_t$
starting with $N_f$ filament seeds of three units attached to the left wall 
(given our choice of three beads filaments as effective seeds, see 
Eq.~(\ref{eq:reaction})) and a set of $N_t-3N_f$ free monomers disposed at 
3D cubic lattice vertices, to optimize initial space homogeneity of the monomer 
density. 
Note that in preparing the equilibium trajectory of any bundle+free monomers 
solution state point by a preliminary explicit growth of the bundle from a 
non-equilibrium initial configuration with filament seeds and free monomers 
only, we automatically tested the robustness of our simulation scheme in 
reproducing in a systematic way the studied network equilibrium state.

The results of the three series of experiments I, II (including IIa and IIb) 
and III ordered by increasing grafting density, are gathered in 
Tables~\ref{tab.3},~\ref{tab.4} and~\ref{tab.5} respectively. 
Within each table, the data are organized such that the total number $N_t$ 
of monomers in the system increases from top to bottom, so that the system 
evolves typically from a subcritical regime up to more and more supercritical 
regimes (note that in experiment series I and IIb, only supercritical regimes 
are considered). 

\begin{table}
\caption{List of simulation experiments performed at $k_{\rm B}T=1$, 
$L=16d$, $l_{\rm p}=250d$ and $\nu=50 u_t^{-1}$, regrouped into four experiment 
series in which only the total number of monomers $N_t$ is changing. 
The transverse area is $A=H^2$, the brush surface density is 
$\sigma_f=N_f/A$, the equilibrium constant $K_0$ (see Eq.~(\ref{eq:K0})) 
is based on different values of $\epsilon_0'$, namely 8.04211 (I, IIb), 
 7.34894 (IIa) and 6.61497 (III).}
\label{tab.1}
\begin{center}
\begin{tabular}{lccccr}
\hline
\hline
Experiment  & H/d & $N_f$ & $N_t$ & $\sigma_f d^2$ & $K_0/d^3$ \\
 &  &  &   &  &   \\
\hline
I & 16 & 32 &  $450,500,525$ & 0.1250  &  78.1397 \\
&  &  &  $550,600$ &  &   \\ 
IIa  & 12 & 32 & $230,300,370$ & 0.2222  & 39.0698  \\
  &  & & $437,500,550$ &   &   \\
IIb  & 12 & 32 & $371,438$ & 0.2222  & 78.1397  \\
  &  & & $501,551$ &   &   \\
III  & 10 & 32 & $240,290,350$ & 0.3200  &  18.7535  \\
  & & & $430,490,540$ &  &   \\
\hline
\hline
\end{tabular}
\end{center}
\end{table}

\subsection{Distribution of filament lengths and free monomer distribution
\label{sec:simul_tech}}

\begin{figure}
\begin{center}
\includegraphics[scale=0.45]{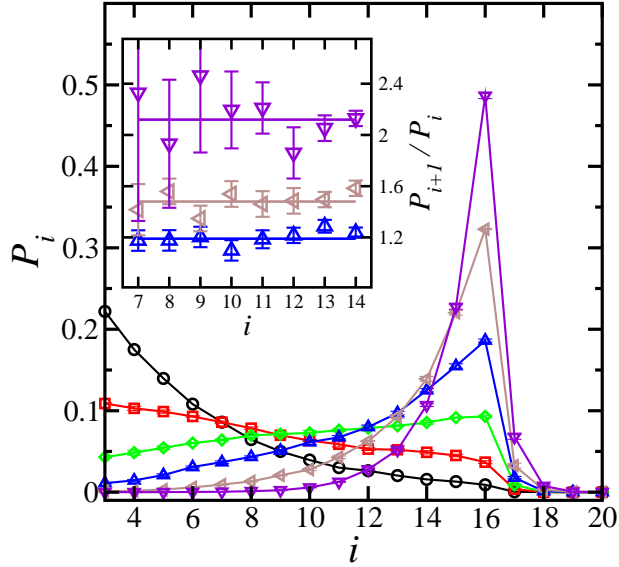}
\caption{Distribution of filament sizes obtained in the simulation series 
IIa of experiments at surface density $\sigma_f d^2=0.222$ for $N_t$ values 
$230$ (circles), $300$ (squares), $370$ (lozenges), 
$437$ (triangles pointing up), 
$500$ (triangles pointing left) and $550$ (triangles pointing down). 
Continuous lines have been drawn for better data visualization. 
Inset: the ratios $P_{i+1}/P_i$ are plotted versus $i$ (for $i<z$) for the 
experiments $N_t=437,500,550$ to which corresponds, in the same order, an 
increasing $\hat{\rho}_1^{\rm eff}$ fitting value indicated by an horizontal 
line (see text). 
}
\label{fig:ee}
\end{center}
\end{figure}
\begin{figure}
\begin{center}
\includegraphics[scale=0.45]{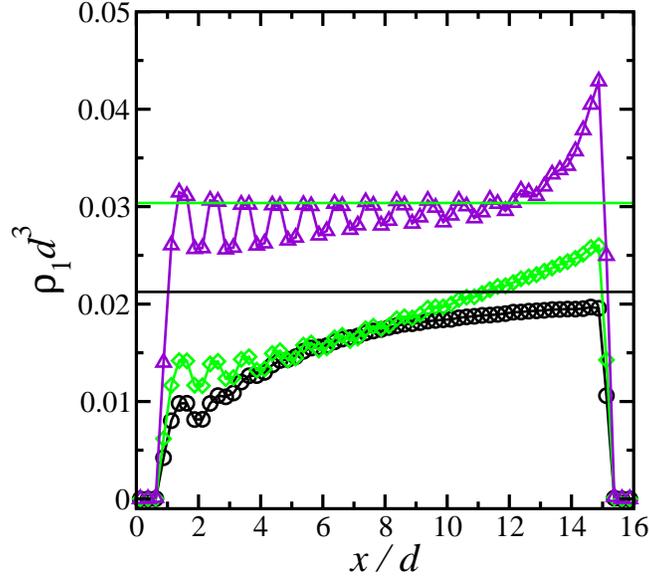}
\caption{Free monomer density as a function of $x/d$ from the anchoring wall 
at ($x=0$) to the obstacle wall at ($x/d=16$) in the simulation series 
IIa of experiments at surface density $\sigma_f d^2=0.2222$, for $N_t$ values 
$230$ (black circles), $370$ (green lozenges), 
$550$ (violet triangles). 
Corresponding values of $\rho_1^{\infty}$ are indicated as a continuous 
horizontal line of same color except for the highest density 
$\rho_1^{\infty}d^3=0.0725$ (for $N_t=550$) lying outside the shown 
density window.}
\label{fig:ff}
\end{center}
\end{figure}
Figures~\ref{fig:ee} and~\ref{fig:ff} show respectively the distributions 
of filament sizes and the corresponding free monomer densities averaged along 
the transverse directions, for the various state points considered at 
intermediate density $\sigma_f d^2=0.2222$. 

We observe in Fig.~\ref{fig:ee}, that all distributions are reasonably smooth 
up to $i=16$ (inclusive).
This is coherent with an independent filament Monte-Carlo prediction of the 
wall factors defined by Eq.~(\ref{eq:wallfac}), using the Hamiltonian $H_N$ 
given by Eq.~(\ref{eq:hamil}) with parameters taken identical to those adopted 
in our MD simulations. 
Given the slightly flexible bonds with fluctuations ($\sigma_d/d=0.05$) and the 
range of the wall potential ($x_{\rm c}/d \cong 1.20$),  the index of the 
largest filament which has essentially no direct interactions with the wall is 
$z=15$. 
The $\alpha_{z+k}$ were obtained by a Metropolis MC sampling of an anchored 
filament of size $z=15$ to which up to six extra bonds are added to probe the 
wall potential energy Boltzmann factor.~\cite{RRMP.13}
We get $\alpha_{16}=0.9245(4)$, $\alpha_{17}=0.11700(2)$, 
$\alpha_{18}=0.01400(4)$, $\alpha_{19}=0.0026(1)$, $\alpha_{20}=0.00069(6)$ 
and $\alpha_{21}=0.00024(2)$,indicating a weak wall perturbation for $i=16$ 
and a strong wall perturbation for larger sizes.
According to Eq.~(\ref{eq:pop}), the size distribution in the ideal case will 
show a clear drop for $i \geq 17$, and this is observed in our MD results also.

The free monomer density $\rho_1(x)$ is now a field variable and we observe that
 it is correlated to the co-volume of the filaments which follow a size
exponential distribution shown in Fig.~\ref{fig:ee}. 
Among the six systems considered at the same temperature and same surface 
density, the first system is clearly subcritical (apparent exponential decay 
of filament size populations) while the three last cases are supercritical 
(apparent exponential increase of filament size populations before wall 
influence). 
The second and third systems are close to the critical regime with a slight 
sub- or supercritical character respectively. 

We found that the size distributions for  $i \leq z=15$ are all very well 
described by an exponential function,
\begin{equation}
P_i \propto \exp{\left[\ln{(\hat{\rho}_1^{\rm eff})}i \right]},
\label{eq:piexp}
\end{equation}
implying according to Eq.~(\ref{eq:equi_const3})
\begin{equation}
P_i/P_{i-1}=\rho_1^{\infty}  f_1^{\infty}\frac{f_{i-1}}{f_i}K_{0} \approx \hat{\rho}_1^{\rm eff}.
\label{eq:rho1eff}
\end{equation}
Eq.~(\ref{eq:rho1eff}), verified in the inset of Fig.~\ref{fig:ee}, suggests 
that the ratio $r_i=f_{i-1}/f_i$ is effectively independent of the filament 
size. 
In Tables~\ref{tab.3},~\ref{tab.4} and~\ref{tab.5} are reported all estimates 
of the effective free monomer density $\hat{\rho}_1^{\rm eff}$, a quantity which
 turns out to provide a measure of the degree of supercriticality of a 
particular state point and which reduces naturally to 
$\hat{\rho}_1=K_0 \rho_1$ in the ideal case when all activity coefficients are 
unity and $\rho_1^{\infty}$ becomes equivalent to the uniform $\rho_1$ value.

As a digression, we now discuss some preliminary data of experiment IIa listed 
separately in Table~\ref{tab.2}. 
They provide a general structural and dynamical information which helps 
appreciating some general trends as $N_t$ increases and also allows us to verify
 that simulation averages deal with well equilibrated systems. 
With $\rho_t$ increasing, the average size of filaments and the correlated 
average height (end-to-end vector projected onto $x$) of the brush 
systematically increases. 
As could be expected, the amplitude of the size fluctuations and their 
characteristic relaxation time $\tau_i$ are largest close to the critical 
regime ($\hat{\rho}_1^{\rm eff}=1$).
We note that the sampling time window of each of the four independent MD 
runs per experiment (whose averages are provided in tables) is at least a 
factor ten larger than the intrinsic relaxation time $\tau_i$. 
This explains our choice to adopt a larger value of the $\nu$ parameter with 
respect to our first study in Ref.~\cite{Caby.12}
The averaged (de)polymerization rates $\langle U \rangle $ and 
$\langle W \rangle$ per filament-end in Table~\ref{tab.2} are averages over all 
filament sizes. 
The same frequency observed for polymerization and depolymerization events 
is an additional indication of equilibrium. 
As $\rho_t$ increases, one first observes an increase of the chemical events 
frequency as a result of the increase of the free monomer density and in 
parallel, an increase of the fraction of filaments able to depolymerize
($i \geq 4$). 
As $\rho_t$ further increases, the overall rates start decreasing because the 
wall presence forbids easy polymerization of the most populated long 
filaments and at the same time the relatively high packing fraction causes a 
decrease of the acceptance rate of the attempted chemical 
events.~\cite{Caby.12}
\begin{table}
\caption{Equilibrium data on a brush of $N_f=32$ filaments pressing 
against a fixed wall at density $\sigma_f d^2=0.2222$ 
(Experiment IIa with $K_0d^{-3}=39.0698$). 
$N_t$ is the total number of monomers in the volume $V/d^3=2304$ which 
also contains $1440$ MPCD solvent particles. 
$\langle i \rangle$ is the average size of a filament expressed in the 
number of monomers, including those which initiate the filament at the 
left wall. 
The two next columns provide data on the filament size fluctuations, namely 
their amplitude and characteristic relaxation time $\tau_i$. 
$\langle X_{i} \rangle$ is the averaged projection of the end-to-end vector 
of the filaments on the normal to the walls ($x$ axis). 
$\langle U \rangle $ and $\langle W \rangle$ are the average (de)polymerization 
rates per free filament end. 
Note that times and frequencies are made dimensionless by using the time 
unit $u_t$ defined in Sec.~\ref{sec:simul_exp} and  unmentioned errors 
are of one unit on the last digit indicated.}
\label{tab.2}
\begin{center}
\begin{tabular}{c c c c c c c c}
\hline
\hline
& & & & & &\\
$N_{\rm t}$&
$\langle i\rangle$&
$\langle \delta i^2\rangle^{1/2}$&
$\tau_i/(10^3 u_t)$ &
$\langle X_{i}/d \rangle$ &
$\langle U u_t \rangle $ &
$\langle W u_t \rangle $ \\
& & & & & &\\
\hline
& & & & & &\\
230&  				
6.16 &  				
3.16&	
5.6 & 
5.12 (1)&    
0.00208&  			
0.00208&  			
\\
300&  				
8.26 &  				
3.91&	
8.2 & 
7.17 (1)&    
0.00236&  			
0.00236&  			
\\
370&  				
10.39 &  					
3.94	&	
7.9 &  
9.26 (1)&    
0.00241&  			
0.00241&  			
\\
437&  				
12.41 &  					
3.41&			
7.3 &   
11.20 (1) &    
0.00233&  			
0.00233&  			
\\
500&  				
14.17 &  					
2.32 &		
4.0 &  
12.92 (2)&    
0.00214&  			
0.00215&  			
\\
550&  				
15.24 &  					
1.38 &		
1.4 &    
13.93 (2)&    
0.00187&  			
0.00187&  			
\\
& & & & & &\\
\hline
\hline
\end{tabular}
\end{center}
\end{table}

\subsection{Data on normal pressure/compressive force and free monomers 
chemical potential 
\label{sec:pres}}

The force exerted by a monomer $i$ located at a close distance $x$ from the 
right wall, with $x=x_{\rm wr}-x_i<x_{\rm c}$ where $x_{\rm c}$ is the wall 
interaction cutoff distance, exerts on the wall a positive force deriving 
from Eq.~(\ref{eq:walpot})
\begin{align}
f(x)=
\frac{9 \sqrt{3}}{2} \epsilon_{\rm w} 
\left[3\left(\frac{\sigma_{\rm w}}{x}\right)^9
-\left(\frac{\sigma_{\rm w}}{x}\right)^3 \right]\frac{1}{x}.
\label{eq:forcew}
\end{align}
The pressure on the right wall can be estimated directly as 
\begin{equation}
p_N=p_{\rm w}^{\rm s}+p_{\rm w}^{\rm fm}+p_{\rm w}^{\rm br}
\label{eq:pressure}
\end{equation}
where $p_{\rm w}^{\rm s}$ originates from average normal linear momentum 
exchange per unit area due to solvent (MPCD) particles reflections. 
The free monomer contribution $p_{\rm w}^{\rm fm}$ and the filament 
contribution $p_{\rm w}^{\rm br}$ are the average of the sum per unit area of 
all contributions of the type in Eq.~(\ref{eq:forcew}) coming respectively 
from free monomers and from filament monomers. 
This wall normal pressure (computed at the wall) is equivalent at equilibrium 
to the $xx$-component of the pressure tensor computed at any point in the 
system. 
We did check this property by computing $p_{xx}(x)$ as a sum of all linear 
momentum transfers per unit area across a virtual surface normal to $x$-axis 
at equally spaced values in the range $0<x<L$. 
We found that wall pressure $p_N$ data and volume average of the $p_{xx}$ 
values lead to same estimates of the homogeneous normal pressure (within error 
bars) and the statistical accuracy is similar.

The free monomer chemical potential was computed according to a Widom-like 
virtual particle expression derived in Supplementary Material,~\ref{sec:apA}
for the reactive canonical ensemble, which can be summarized as follows:
\begin{align}
\beta \mu_1
& \equiv \ln {\frac{\Lambda^3}{d^3}}+\beta \mu_1^{*}, \label{eq:final1} \\
\beta \mu_1^{*}&=
-\ln{\left[
\frac{V}{d^3}
\left\langle 
\frac{1}{(N_1+1)} 
\int_0^1 {\rm d}s_{\rm ex}^3 
\exp{[-\beta \Delta U_{\rm ex}]}
\right\rangle \right]},
\nonumber \\
 &=-\ln{
\left[
\frac{V}{d^3}
\left\langle  
\frac{E}{(N_1+1)} 
\right\rangle 
\right]},
\label{eq:final}
\end{align}
where $\langle \cdots\rangle$ represents a reactive canonical ensemble average. 
Within the averaged quantity, $N_1$ is the instantaneous number of free 
monomers in the microscopic configuration and $E$ represents the triple 
integral over $s_{\rm ex}$ which must be estimated, for a given microscopic 
configuration of the system, by a distinct MC sampling of three independent 
random numbers $s_x,s_y$ and $s_z$ in an homogeneous distribution $[0,1]$. 
As indicated in Eq.~(\ref{eq:final}), $E$ requires the average of the 
exponential factor of the additional potential energy 
$\Delta U_{\rm ex}$ due to the interactions between an extra virtual 
particle located at random location $x=s_x L, y=s_y H$ and $z=s_z H$ and both 
the wall and the $N_t$ monomers of the system. 

In Tables~\ref{tab.3},~\ref{tab.4} and~\ref{tab.5}, we report for all three 
series of experiments the averaged wall pressure contributions 
$p_{\rm w}^{\rm fm}$ and $p_{\rm w}^{\rm br}$ defined in 
Eq.~(\ref{eq:pressure}) and the estimates of the average of 
$\langle E/(N_1+1)\rangle $ and corresponding values of 
$\beta \mu_1^{*}$, according to Eq.~(\ref{eq:final}). 
The estimate of $\langle E/(N_1+1)\rangle$ is based on $200$ independent 
extra particle insertions for each treated microscopic configuration, 
the latter being those saved along the trajectory at equally spaced time 
intervals of $100 u_t$ (so a total of $4\times10^3$ configurations per 
experiment).

In the following, we need to know the thermodynamic properties of the free 
monomer solution which would be in equilibrium with the simulated bundle/free 
monomers system and we will denote these reservoir equilibrium properties, 
as already done in Sec. \ref{sec:thermo}, with an $\infty$ superscript. 
We checked that the virial expansion of the pressure $p^{\infty}$ and the 
chemical potential $\mu_1^{*\infty}$ up to the second virial coefficient 
is sufficiently precise for the relevant concentration regime, and so we used
\begin{align}
\frac{p^{\infty}}{k_{\rm B}T}
&\cong \rho_1^{\infty}(1+B_2 \rho_1^{\infty}),
\label{eq:virial}\\
\beta \mu_1^{*\infty} &\cong \ln{\left(\rho_1^{\infty}d^3 \right)}
+2 B_2 \rho_1^{\infty}, \label{eq:mu1**}
\end{align}
where $B_2/d^3=1.722824$ at $k_{\rm B}T^{\infty}=1$, assuming the shifted LJ 
potential with cutoff at minimum (see Sec.~\ref{sec:simul_exp} or see Eq.~(2) 
in Ref.~\cite{Caby.12}).
Imposing $T^{\infty}=T$ and $\mu_1^{*\infty}=\mu_1^{*}(N_t,N_f,A,L,T)$
where $\mu_1^{*}$ is known from Eq.~(\ref{eq:final1}), we get from 
Eqs.~(\ref{eq:virial}) and~(\ref{eq:mu1**}) the pressure and density of the free 
monomer fluid reservoir.

Using the second equality of Eq.~(\ref{eq:rho1eff}), the $i$ independent 
ratio $r=f_{i-1}/f_i$ can be estimated from the reduced density 
$\hat{\rho}_1^{\rm eff}$ (as reported in Tables) while the free monomer 
chemical potential $\mu_1^{*}$ is obtained using the connection 
$d^3 \rho_1^{\infty} f_1^{\infty}=\exp{(\beta \mu_1^{*})}$ which follows from 
Eqs.~(\ref{eq:mu1}) and~(\ref{eq:final1}).

For the evaluation of the osmotic pressure $\Pi$ in Eq.~(\ref{eq:pi}), 
we need to subtract from $p_N$ the common solvent pressure and the pressure 
$p^{\infty}(\mu_1^{*},T)$ evaluated by combining Eqs.~(\ref{eq:virial}) and
~(\ref{eq:mu1**}). 
The average normal compressive force per filament follows as 
$f_N=\pi/\sigma_f$, as reported in Tables~\ref{tab.3},~\ref{tab.4} 
and~\ref{tab.5}. 
Concerning the free monomer density, the same Tables mention the space 
averaged  value $\langle \rho_1 \rangle $ within the bundle which is 
systematically smaller than the free monomer density $\rho_1^{\infty}$ of the 
reservoir solution in equilibrium with it. 
In Fig.~\ref{fig:ff}, for each state point shown, we indicate by an 
horizontal line the $\rho_1^{\infty}$ value corresponding to the 
$\rho_1(x)$ profile.

\begin{table*}
\caption{Equilibrium data on a brush of $N_f=32$ filaments pressing against a 
fixed wall at density $\sigma_f d^2=0.125$ (Experiment I). 
$N_t$ is the total number of monomers in the volume $V/d^3=4096$ which also 
contains $2560$ MPCD solvent particles. 
$\langle\rho_1\rangle$ is the average free monomer density in the total volume 
$V$. 
The quantity $\langle E/(N_1+1)\rangle $ is the average expression in the 
argument of the logarithm term used in Widom-like formula Eq.~(\ref{eq:final}), 
which then leads to $\beta \mu_1^{*}$ reported in next column. 
The two next columns provide the reduced free monomer concentration defined 
as $\hat{\rho}_1=\exp{(\beta \mu_1^{*})} K_0$ (with $K_0/d^3=78.13968$  
in the present case) and the effective reduced density  
$\hat{\rho}_1^{\rm eff}$ obtained by an exponential fitting of the 
$P_i$ simulation data. 
The next column provides an estimate of the apparently $i$ independent ratio 
$r=f_{i-1}/f_i$ evaluated as the ratio $\hat{\rho}_1^{\rm eff}/\hat{\rho}_1$. 
$p_{\rm w}^{\rm br}$ and $p_{\rm w}^{\rm fm}$ are respectively the 
contributions of the brush and the free monomers to the pressure exerted on the 
right wall, $p^{\infty}$ is the pressure in the pure monomer solution in 
chemical equilibrium with the brush. 
The last column provides the reduced force per filament computed as 
$f_N=(p_{\rm w}^{\rm br}+p_{\rm w}^{\rm fm}-p^{\infty}) \sigma_f^{-1}$. 
Note that the limit for buckling for an independent filament is 
$\hat{\rho}_1^{\rm b}=\exp{(l_{\rm p} d/L^2)}=2.66$, so that the 
$N=600$ case needs to be considered with care as the effective 
$\hat{\rho}_1^{\rm eff} =2.6 \cong \hat{\rho}_1^{\rm b}$.}
\label{tab.3}
\begin{center}
\begin{tabular}{c c c c c c c c c c c c c c}
\hline
\hline
& & & & & & & & & & & &\\
$N_{\rm t}$&
$\langle \rho_1 d^3 \rangle$&
$\langle E/(N_1+1)\rangle$&
$\beta \mu_1^{*} $&
$\hat{\rho}_1$&
$\hat{\rho}_1^{\rm eff}$ &
$r$&
$\beta p_{\rm w}^{\rm br} d^3$ &
$\beta p_{\rm w}^{\rm fm} d^3$&
$\beta p^{\infty} d^3$&
$\beta f_N d$\\
& & & & & & & & & & & &\\
\hline
& & & & & & & & & & & &\\
450&  				
0.01109(3) & 		
0.0140(1) &  			
-4.049(7) &			
1.36(1)		&	
1.21(1)  &  			
0.89(2)  &      
0.0261(7)  &	 
0.0157(1)  &  			
0.01689    &   
0.200(6)  &
\\
500&  				
0.0125(3)& 		
0.0119(1) &  			
-3.887(8) &			
1.60(1) &  			
1.46(1)  &  			
0.91(2)  &      
0.0439(6)    &	 
0.0182(4)   &  			
0.01989 &   
0.338(6) &
\\%
525&  				
0.0141(2)& 		
0.01025(8) &  			
-3.737(8) &			
1.86(1) &  			
1.59(1)  &  			
0.85(2) &      
0.0597(6)   &	 
0.0206(4)   &  			
0.02292 &   
0.459(6) &
\\%
550&  				
0.0164(2) & 		
0.00862(6) &  			
-3.564(7) &			
2.21(2) &  			
1.86(1) &  			
0.84(2) &      
0.073(1)   &	 
0.0235(1) &  			
0.02705  &   
0.557(8)  &
\\
600&  				
0.0227(1)& 		
0.00591(3) &  			
-3.187(5) &			
3.23(2) &  			
2.64(1)  &  			
0.82(2) &      
0.116(3)  &	 
0.0320(3) &  			
0.03875 &   
0.87(2) &
\\
& & & & & & & & & & & &\\
\hline
\hline
\end{tabular}
\end{center}
\end{table*}

\begin{table*}
\caption{Equilibrium data on a brush of $N_f=32$ filaments pressing against a 
fixed wall at density $\sigma_f d^2=0.2222$ (Experiment II). 
The first set of data coined as experiment IIa and the second set of data 
coined as experiment IIb differ only by a different choice of $\epsilon_0'$ 
(see Table~\ref{tab.1}) leading to different $K_0$ values indicated in the 
extra second column. 
$N_t$ is the total number of monomers in the volume $V/d^3=2304$ which also 
contains $1440$ MPCD solvent particles. 
See caption of similar Table~\ref{tab.3} for explanations on the nature of the 
shown data and for the last caution sentence which applies here also for the 
$N=551$ case in Experiment IIb.}
\label{tab.4}
\begin{center}
\begin{tabular}{c c c c c c c c c c c c c c c}
\hline
\hline
& & & & & & & & & & & & &\\
$N_{\rm t}$&
$K_0 d^{-3}$  &
$\langle \rho_1 d^3\rangle$&
$\langle E/(N_1+1)\rangle$&
$\beta \mu_1^{*} $&
$\hat{\rho}_1$&
$\hat{\rho}_1^{\rm eff}$ &
$r$&
$\beta p_{\rm w}^{\rm br} d^3$ &
$\beta p_{\rm w}^{\rm fm} d^3$&
$\beta p^{\infty} d^3$&
$\beta f_N d$\\
& & & & & & & & & & & & &\\
\hline
& & & & & & & & & & & & &\\
230&  				
39.0698&  				
0.0143(1)& 		
0.0205(2) &  			
- 3.86(1) &			
0.827(8) &  			
0.79(2) &  			
0.96  &   
0.0015(1)  	&		
0.0203(2) &  			
0.0205 &
0.006(1)&
\\
300&  				
39.0698&  				
0.0155 (1)& 		
0.0169(1) &  			
-3.662(5)&  			
1.003(6) &  			
0.92(2) &  			
0.92  &  
0.0079(4)  	&		
0.0239(2)&  			
0.0246&
0.032(2) &
\\
370&  				
39.0698&  				
0.0163 (1)& 		
0.0143(1) &  			
-3.495(5) &  			
1.186(8) &  			
1.04(2) &  			
0.88  &   
0.021(1)  	&		
0.0274(3) &  			
0.0290 &
0.088(5)&
\\
437&  				
39.0698&  				
0.0174 (1)& 		
0.0118(1)&  			
-3.303(7) &  			
1.44(1) &  			
1.19(2) &  			
0.83  &   
0.046(1)	&		
0.0316(2) &  			
0.0348 &
0.193(5) &
\\
500&  				
39.0698&  				
0.0203 (2)& 		
0.00890(7)&  			
-3.021(7)&  			
1.91(1)&  			
1.48(2) &  			
0.77  &   
0.087(3)	&		
0.0375(3)&  			
0.045 &
0.36(1) &
\\
550&  				
39.0698&  				
0.0270 (2)& 		
0.00599(3) &  			
-2.625(5)&  			
2.83(1)&  			
2.12(2) &  			
0.75  &   
0.156(2)	&		
0.0485(5) &  			
0.0652 &
0.63(1) &
\\
& & & & & & & & & & & & &\\
371&  				
78.13968&  				
0.0081 (1)& 		
0.0286(3) &  			
-4.189(1) &  			
1.19(1) &  			
1.09(2) &  			
0.92  &   
0.028(2)  	&		
0.0139(2) &  			
0.0148 &
0.122(9)&
\\
438&  				
78.13968&  				
0.00884 (3)& 		
0.0232(3)&  			
-3.98(1) &  			
1.46(2) &  			
1.27(2) &  			
0.87  &   
0.058(1)	&		
0.0162(1) &  			
0.0181 &
0.252(7) &
\\
470&  				
78.13968&  				
0.0096 (1)& 		
0.0199(2)&  			
-3.83(1) &  			
1.70(2) &  			
1.44(2) &  			
0.85  &   
0.077(2)	&		
0.0179(2) &  			
0.0210 &
0.33(1) &
\\
501&  				
78.13968&  				
0.0116(2)& 		
0.0156(2)&  			
-3.58(1)&  			
2.18(2)&  			
1.74(2) &  			
0.80  &   
0.120(1)	&		
0.0212(5)&  			
0.0267 &
0.513(5) &
\\
551&  				
78.13968&  				
0.0198 (3)& 		
0.00820(7) &  			
-2.94(1)&  			
4.13(4)&  			
2.98(2) &  			
0.72  &   
0.233(3)	&		
0.0333(3) &  			
0.0488 &
0.98(1) &\\
& & & & & & & & & & & & &\\
\hline
\hline
\end{tabular}
\end{center}
\end{table*}

\begin{table*}
\caption{Equilibrium data on a brush of $N_f=32$ filaments pressing against 
a fixed wall at density $\sigma_f d^2=0.32$ (Experiment III). 
$N_t$ is the total number of monomers in the volume $V/d^3=1600$ which also 
contains $1000$ MPCD solvent particles. 
See caption of similar Table~\ref{tab.3} for explanations on the nature of 
the shown data. 
Note that $K_0/d^3=18.75352$ in the present case.}
\label{tab.5}
\begin{center}
\begin{tabular}{c c c c c c c c c c c c c c}
\hline
\hline
& & & & & & & & & & & &\\
$N_{\rm t}$&
$\langle \rho_1 d^3\rangle$&
$\langle E/(N_1+1)\rangle$&
$\beta \mu_1^{*} $&
$\hat{\rho}_1$&
$\hat{\rho}_1^{\rm eff}$ &
$r$&
$\beta p_{\rm w}^{\rm br} d^3$ &
$\beta p_{\rm w}^{\rm fm} d^3$&
$\beta p^{\infty} d^3$&
$\beta f_N d$\\
& & & & & & & & & & & &\\
\hline
& & & & & & & & & & & &\\
240&  				
0.0274(1)& 		
0.0130(1) &  			
-3.035(7) &			
0.902(7) &  			
0.79(2) &  			
0.88 &      
0.0029(1)   &	 
0.0441(2) &  			
0.04461 &
0.0076(4) &
\\
290&  				
0.0282(1)& 		
0.01113(6)&  			
-2.879(5) &			
1.053(6) &  			
0.87(2) &  			
0.83 &   
0.0091(4) 	&		
0.0503(3)&  			
0.05159 &
0.025(1) &
\\
350&  				
0.0281(1) & 		
0.00947(6) &  			
-2.718(6) &  			
1.238(8) &  			
1.00(2)  &  			
0.81 &    
0.0233(7)  	&		
0.0566(2) &  			
0.05984 &
0.063(2)&
\\
430&  				
0.0282(2)& 		
0.00732(6)&  			
-2.461(8) &  			
1.60(1) &  			
1.17(2) &  			
0.73  &     
0.0655(7)	&		
0.0670(4) &  			
0.07548 &
0.178(3) &
\\
490&  				
0.0295(1)& 		
0.00562(4)&  			
-2.196(7) &  			
2.09(1) &  			
1.41(2) &  			
0.67 &   
0.121(1)	&		
0.0756(5)&  			
0.09539 &
0.315(3) &
\\
540&  				
0.0369(2)& 		
0.00370(3)&  			
-1.778(8) &  			
3.17(3) &  			
1.98(2) &  			
0.62 &     
0.2060(5)  &		
0.0933(7) &  			
0.1365 &
0.509(3) &
\\
& & & & & & & & & & & &\\
\hline
\hline
\end{tabular}
\end{center}
\end{table*}

\begin{figure}
\begin{center}
\includegraphics[scale=0.45]{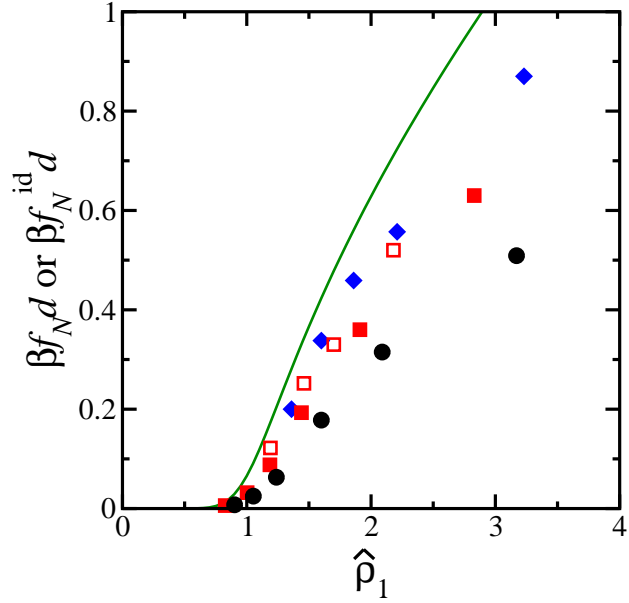}
\caption{Osmotic force per filament $f_N$ as a function of 
$\hat{\rho}_1=K_0 \rho_1^{\infty} f_1^{\infty}$ for 
simulation data corresponding to various surface densities 
$\sigma_f d^2=0.125$ (Experiment I, blue lozenges), $0.222$ 
(red filled squares for Experiment IIa and red open squares for Experiment IIb), 
and $0.320$ (Experiment III, black circles) compared to the ideal 
solution prediction of the force per filament $f_N^{\rm id}$ as a 
function of $\hat{\rho}_1$ based on the 
filament-wall microscopic model used in the bundle simulations. 
}
\label{fig:cc}
\end{center}
\end{figure}

In Fig.~\ref{fig:cc}, we plot $f_N$ as a function of 
$\hat{\rho}_1 =K_0 \rho_1^{\infty} f_1^{\infty}$ that we compare with the ideal 
solution prediction $f_N^{\rm id}$ plotted against 
$\hat{\rho}_1=\rho_1 K_0$. The data $f_N^{\rm id}$ were computed according to Eq.~(\ref{eq:fn}) using distributions (\ref{eq:popi},\ref{eq:pop},\ref{eq:ddd})
for $L/d=16$, $z=15$ and $k_{\rm B}T=1$ on the basis of the $\alpha$'s and their $L$ derivatives obtained by the single filament Monte-Carlo simulations 
~\cite{RRMP.13} already discussed in Sec.~\ref{sec:simul_tech}.

Experiments I and IIb, which have the same ideal solution chemical 
equilibrium constant, present for the same reservoir free monomers state 
point, i.e. at the same $\hat{\rho}_1$ value, a weak but systematic decrease 
of the force $f_N$ with increasing density. 
If we compare the force $f_N$ between experiments IIa and IIb which correspond 
to the same grafting density while $K_0$ is twice larger in the IIb case, 
we observe that for the same $\hat{\rho}_1$ which corresponds to a much lower 
free monomer density $\rho_1^{\infty}$ in IIb (would be twice smaller if system 
was ideal), the force is significantly larger in IIb than in IIa. 
If we now compare results of experiments I, IIa and III in that sequence, 
the points are shifted more and more away from the single independent 
filament prediction, as an increasing grafting density and a decreasing 
equilibrium constant $K_0$ tend both to a reduction of the force.

\begin{figure}
\begin{center}
\includegraphics[scale=0.45]{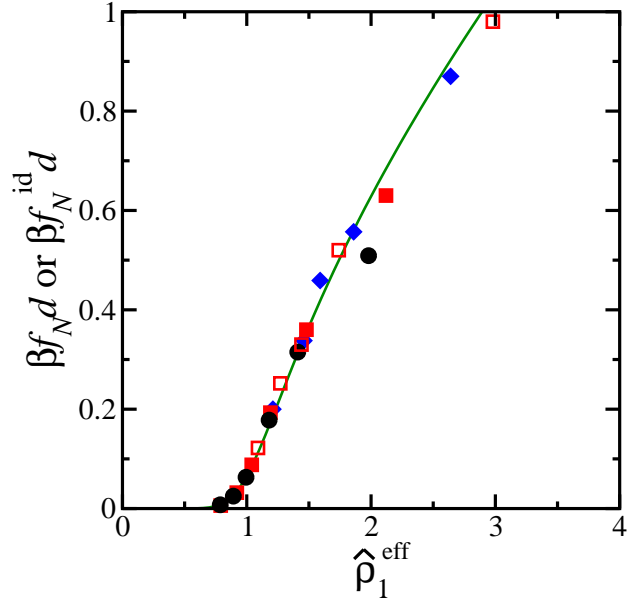}
\caption{Osmotic force per filament $f_N$ as a function of 
$\hat{\rho}_1^{\rm eff}$ for simulation data corresponding to various surface 
densities $\sigma_f d^2=0.125$ (Experiment I, blue lozenges), 
$0.222$ (red filled squares for Experiment IIa and red
open squares for Experiment IIb), 
and $0.320$ (Experiment III, black circles) compared to the ideal solution 
prediction of the force per filament $f_N^{\rm id}$ as a function of 
$\hat{\rho}_1$ based on the filament-wall microscopic model used in the bundle 
simulations.
}
\label{fig:dd}
\end{center}
\end{figure}

In Fig.~\ref{fig:dd}, we plot the same values of $f_N$ as a function of 
$\hat{\rho}_1^{\rm eff} 
=\hat{\rho}_1 r
= K_0 \rho_1^{\infty} f_1^{\infty}f_{i-1}/f_i $ 
and observe that all points align along the same ideal solution curve given by 
Eq.~(\ref{eq:fn}) already shown in Fig.~\ref{fig:cc}. 
This is the central result of this paper as it shows that the osmotic force 
per filament 
\begin{equation}
f_N(N_t,N_f,A,L,T)=\phi(L,T,\hat{\rho}_1^{\rm eff})
\label{eq:univ}
\end{equation}
where the ideal solution function $f_N^{\rm id}=\phi(L,T,\hat{\rho}_1)$ is 
given by Eq.~(\ref{eq:fn}) and where in the non ideal case, all interactions 
are taken into account by a renormalization of the effective reduced free 
monomer density leading to $\hat{\rho}_1^{\rm eff}$. 
The comparison of Eq.~(\ref{eq:univ}) and Eq.~(\ref{eq:fn}) implies, assuming an
 effective state point independence of the mean wall force $\bar{f}_{z+k}(L)$ on
 a filament of size $z+k$, that
\begin{equation}
P_i(N_t,N_f,A,L,T) \cong P_i^{\rm id}(L,T,\hat{\rho}_1^{\rm eff})
\label{eq:univ1}
\end{equation}
for all filament sizes $(i=3,z^{*})$, where the ideal solution number density 
expressions are given by Eqs.~(\ref{eq:popi}),~(\ref{eq:pop}) 
and~(\ref{eq:ddd}). 
We checked property Eq.~(\ref{eq:univ1}) which was found to be reasonably 
verified, taken into account the relatively large statistical uncertainties 
on the densities for $i>z$.

The renormalized reduced free monomer density $\hat{\rho}_1^{\rm eff}$ 
involves on one hand a pure solvent property, the product 
$f_1^{\infty}\rho_1^{\infty}$ directly linked 
by Eqs.~(\ref{eq:mu1}) and~(\ref{eq:final1}) to the chemical potential 
$\mu_1^{*}$ and on the other hand the effectively $i$-independent ratio 
$r_i=f_{i-1}/f_i \approx r$ which is basically a measure of the free energy 
required to grow reversibly by one unit a filament of size $i-1$ within the 
bundle, at the state point considered. We have made a detailed analysis of the 
origin of the cost of free energy using Widom like formula
derived originally to compute the incremental excess chemical potential of 
polymers $\mu_i^{\rm exc}-\mu_{i-1}^{\rm exc}$ in a polydisperse 
sample.~\cite{Kumar.91}
The outcome of our analysis is now briefly summarized. 
First, we confirm that these independent $r_i$ estimates are again found to be 
$i$ independent at any state point where they reproduce, within a few percents, 
the ratio of activity coefficients given in Tables~\ref{tab.3},~\ref{tab.4} 
and~\ref{tab.5}, based on the characteristic lengths of the exponential size 
distribution of the filaments at each state point (Eq.(\ref{eq:rho1eff}).
The free energy cost involves the reversible work of growing the filament 
against repulsive forces coming both from the other filaments and from the 
free monomers within the solution. 
When looking at the results within any experiment series (I, IIa, IIb or III), 
we find that, with $\hat{\rho}_1$ increasing (see data in Tables), the observed 
systematic decrease of $r$ can be traced as originating from a combination of 
both types of repulsive interactions. 
However, when comparing data from experiments I and IIb which have different 
grafting densities but the same equilibrium constant $K_0$, we find that for 
two state points taken at the same $\hat{\rho}_1$ ($\mu_1$) value, the global 
$r$ factor is lower for the denser brush (exp IIb) solely, as a result of the 
more important interfilament interactions in the denser denser brush case. 
This can be observed by comparing cases $N_t=550$ in experiment I
(Table~\ref{tab.3}) and $N_t=501$ in experiment IIb (Table~\ref{tab.4}) for 
which we have the same $\hat{\rho}_1 \cong 2.2$. 
When comparing experiments IIa and IIb with the same grafting density but with 
equilibrium constants $K_0$ differing by a factor two, one observes for a 
given similar value of $\hat{\rho}_1$ in both experiments that to the highest 
$K_0$ value corresponds the largest $r$ value solely as a result of the lower 
free monomers density, as the contribution to $r$ from inter-filament 
interactions remain similar at the same grafting density. 
This can be seen by comparing in Table~\ref{tab.4}, the case $N_t=437$ of 
experiment IIa and the case $N_t=438$ of experiment IIb for which 
$\hat{\rho}_1 \approx 1.45$. 


The above discussion concerns a large set of simulations considered for 
different grafting densities $\sigma_f$, different equilibrium constants 
$K_0$ and different free monomer chemical potential $\mu_1$. 
These simulations deal however with a single specific box length $L/d=16$, 
a single temperature $T$, a single persistence length $l_{\rm p}$ and a 
specific filament-wall model $U_{\rm w}(r)$ and it must be added that the 
filament arrangement within the bundle is also specific as all seeds 
(say one of the two first monomers anchored in the left wall) are located 
within a strictly planar wall interface. 
In that context, the functional form of $f_N^{\rm id}=\phi(L,T,\hat{\rho}_1)$ shown in Fig.~\ref{fig:cc}
and hence of $f_N=\phi(L,T,\hat{\rho}_1^{eff})$ in the non ideal conditions 
of the simulations, should be highly specific to the 
choice of all these parameters. 

\begin{figure}
\begin{center}
\includegraphics[scale=0.45]{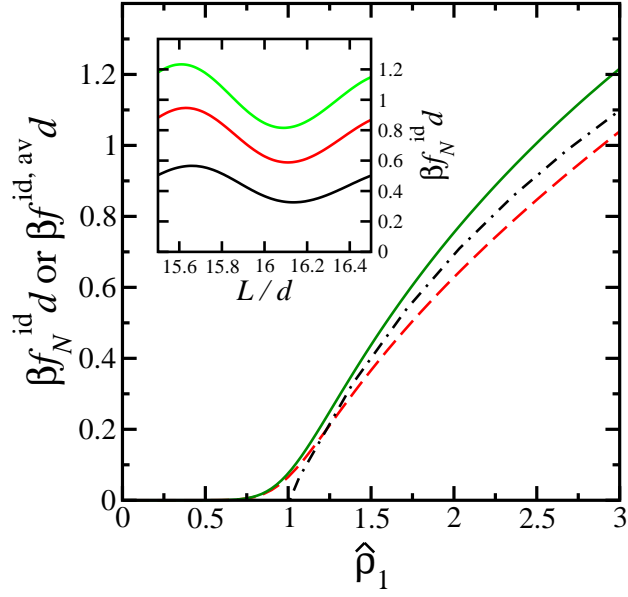}
\caption{Local reduced force $\beta f_N^{\rm id} d$ (red dashed curve) and 
local reduced average force $\beta f_{\rm N}^{\rm id,av} d$ (green continuous 
curve) exerted by the right wall located at gap distance $L/d=16$ on one 
filament anchored normally at the left wall and growing towards the right, 
as a function of $\hat{\rho}_1$, for the filament model (with $l_{\rm p}=250d$) 
and for the filament-wall interaction model which are used in bundle 
simulations. 
The black dash-dotted curve shows $\beta f_N^{\rm id} d = \ln \hat\rho_1$ 
predicted by Eq.~(\ref{eq:force}).
Inset: same local reduced force $\beta f_N^{\rm id} d$ for a right wall 
position varying between $L/d=15.5$ and $L/d=16.5$. 
The three curves (from bottom upwards) correspond to free monomer reduced 
number densities $\hat{\rho}_{1}=1.5$, $\hat{\rho}_{1}=2.0$ and 
$\hat{\rho}_{1}=2.5$. 
}
\label{fig:locfor1}
\end{center}
\end{figure}

Reasoning on the $L$ dependence of the force in the ideal bundle case is 
relatively straightforward~\cite{RRMP.13}. 
Explicit calculations of $f_N^{\rm id}$ show pronounced local fluctuations of 
period $d$ for $L$ variations within the monomer size, the general shape of the 
fluctuations repeating with some damping effect as $L$ increases over a few $d$ 
distances (remaining within the regime $L\ll l_{\rm p}$). 
The shape of these fluctuations is illustrated in the inset of 
Fig.~\ref{fig:locfor1} for the model Hamiltonian used for our simulations. 
By averaging Eq.~(\ref{eq:fn}) at a given $\hat{\rho}_1$ over one monomer size 
interval $[L-d/2<L'<L+d/2]$), one gets trivially
\begin{align}
\beta f_N^{\rm id,av}(L,\hat{\rho}_1)&= 
\frac{1}{d} 
\ln{\left[\frac{D(L+0.5d,\hat{\rho}_1)}{D(L-0.5d,\hat{\rho}_1)}\right]}
\label{eq:forfor}
\end{align}
which can be computed using Eq.~(\ref{eq:ddd}). 
In Fig.~\ref{fig:locfor1}, this average function 
$\beta f_N^{\rm id,av}(L,\hat{\rho}_1)$ for $L/d=16$ is shown together with the 
specific $L/d=16$ expression of $f_N^{\rm id}=\phi(L,T,\hat{\rho}_1)$ already 
shown in the two previous figures. 
Moreover, for comparison, the ratchet model force Eq.~(\ref{eq:force}) is also 
shown for $\hat{\rho}_1>1$. 
It turns out that the average force per filament (in the ideal bundle case) is 
not quantitatively very different from Eq.~(\ref{eq:force}). 
The distinction between both expressions can be analyzed~\cite{RRMP.13} by 
noting that after isolating a $\hat{\rho}_1$ term in the fraction of the 
right hand side of Eq.~(\ref{eq:forfor}), it remains a ratio of two similar 
expressions which are related by the property that any $\alpha_i(L+d/2)$ term 
in the numerator of the fraction is close to a corresponding 
$\alpha_{i-1}(L-d/2)$ term in the denominator of the fraction as they 
correspond to the same universal function 
$\tilde{Z}(\tilde{\eta})$~\cite{Frey.06} computed for two reduced compression 
variables $\tilde{\eta}$ (Eq.~(\ref{eq:etatilde})) differing in relative terms 
by $2d/L_{{\rm c}i}$. 
The universality observed in terms of $\hat{\rho}_1^{\rm eff}$ for the 
particular investigated value of $L/d=16$ in Fig.~\ref{fig:dd} and the 
observation of the weak ($L,l_{\rm p}$) dependence of $f_N^{\rm id,av}$ suggest 
that non ideal systems should be characterized by an average reduced force 
$\beta f_N^{\rm av} d \approx \ln{\hat{\rho}_1^{\rm eff}}$. 
We will come back on this issue in Section~\ref{sec:disc}.

\subsection{Filament size kinetics
\label{sec:kin}}

\begin{figure}
\begin{center}
\includegraphics[scale=0.45]{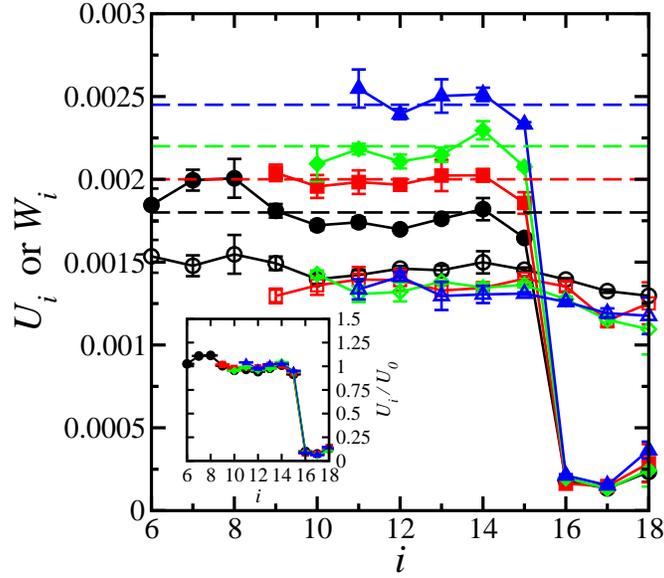}
\caption{
Polymerization rate $U_i$ and depolymerization rate $W_i$ (in units $u_t^{-1}$) 
for filaments of various sizes $i$ in the series of experiments I, 
$N_t=450$ (circles), $500$ (squares) and $525$ (lozenges) and $550$ (triangles). 
Only points known with reasonable statistics (corresponding to significant 
$P_i$ values) are indicated, with filled symbols for $U_i$ 
and empty symbols for $W_i$. 
The error bars being set to one $\sigma$ (estimated from four independent 
runs per experiment). 
Continuous lines are shown to facilitate data observation. 
Estimated bulk constant values $U_0$ (see text) are indicated by horizontal 
dashed lines.
The inset shows the polymerization rates for all values of $i$ in their 
rescaled form $U_i /U_0$.
}
\label{fig:ee2b}
\end{center}
\end{figure}

During all the equilibrium simulations, the number of successful polymerization 
and depolymerization events has been recorded. 
The polymerization and depolymerization rates per active end, 
respectively $U_i$ and $W_i$, for size $i$ filaments are shown in 
Fig.~\ref{fig:ee2b} for experiments of series I, namely those at the lowest 
surface grafting density. 
Only values relative to filament sizes present in significant amount are shown 
in the figure. 
We remind here some qualitative features of the chemical step 
algorithm, all details being found in the original Ref.~\cite{Caby.12}
To get a polymerization step, a free monomer must be present for some time 
at direct proximity of an active end of a filament. 
As long as the monomer is at proximity, with some probability per unit 
time, a polymerization attempt can be tried. 
This attempt consists in transferring spatially this monomer to a tiny 
region of space where the monomer fits the intramolecular structure, i.e. 
is located at a distance close to $d$ from the last monomer of the filament and 
forms a virtual relatively straight trimer with the two last monomers of the 
filament. 
This topological and spatial move is then the object of an 
acceptance/rejection sampling linked to the possible overlap of the displaced 
monomer with the other monomers and/or with the wall. 
For the reverse case, any end-monomer of a filament is essentially all the 
time available for a depolymerization step. With some probability per unit time, an attempt is made to detach and relocate instantaneously the freed 
monomer at proximity of the new filament end. This depolymerization trial step 
is then subjected to an acceptance/rejection on the basis of possible 
overlap with the other monomers and with the wall.
Imposed algorithm micro-reversibility demands that
\begin{equation}
P_i U_i=P_{i+1} W_{i+1}
\label{eq:micro}
\end{equation}
and this property is found to be systematically satisfied in all cases, 
showing simply the correct implementation of our algorithm.~\cite{Caby.12}
The property in Eq.~(\ref{eq:micro}) combined with the observed exponential 
profile of $P_i$ following Eq.~(\ref{eq:piexp}), implies that 
$U_i/W_{i+1}\approx \hat{\rho}_1^{\rm eff}$ in the regime $i<z=15$. 

Figure~\ref{fig:ee2b} shows that in experiment series I, for filament sizes 
($i \leq z=15$) for which direct interaction with the wall is not possible, 
the polymerizing rates $U_i$ (for $i \leq (z-1)=14$) and depolymerizing rates 
for $W_i$ ($i \leq z=15$) are reasonably constant, and we denote by $U_0$ or 
$W_0$ the average values. 
For the different state points, we get for the pairs of (wall) unperturbed 
rates ($U_0 u_t,W_0 u_t$) the values 
$0.0018/0.00148\: (N_t=450)$, 
$0.0020/0.00136 \:(N_t=500)$, 
$0.0022/0.00135 \:(N_t=520)$ and 
$0.00245/0.00133 \:(N_t=550)$. 
As expected from the combination of Eqs.~(\ref{eq:micro}) and~(\ref{eq:piexp}), 
the ratio $U_0/W_0=\hat{\rho}_1^{\rm eff}$  and indeed the values of this ratio 
($1.22$ ($N_t=450$), 
$1.47$ ($N_t=500$), 
$1.63$ ($N_t=525$) and $1.84$ ($N_t=550$)) 
are always compatible with the $\hat{\rho}_1^{\rm eff}$ estimates reported in 
Table~\ref{tab.3}.

Within the series of experiments where only $N_t$ changes from one state point 
to the other, while all other parameters including in particular the equilibrium 
constant $K_0$ are the same, one notes a small decrease of $W_0$ with $N_t$ 
increasing as a result of a decrease in the depolymerization step acceptance 
probability due to increasing volume fraction. The fast 
increase of $U_0$ with $N_t$ is related to the first order character, 
for a given filament end, of the polymerizing reaction as suggested by the 
systematic increase of $\langle\rho_1 \rangle$ with increasing $N_t$
(see Table~\ref{tab.3}). 

All the curves in Fig.~\ref{fig:ee2b} show the same trend for the evolution 
of the (de)polymerization rates as the wall is approached. 
One observes that $W_i$, involving the detachment and re-positioning of the 
freed monomer at proximity of the new filament end, is only slightly affected 
by the presence of the wall repulsive potential. 
The weak decrease of $W_i$ with respect to $W_0$ for $i>z$ must be related to 
a decrease of the acceptance probability of the attempted depolymerizing step, 
as a result of some additional overlap from the wall when the random relocation 
of the freed monomer happens along the end-filament longitudinal direction 
towards the wall.
On the contrary, the polymerization step is strongly affected by the wall 
presence, as the result of the combination of two effects tending to lower 
$U_i$. 
The attempt probability must be lowered by the wall presence as the obstacle 
produces a decrease of the space available to a free monomer to get close to 
the active end and so only a lateral approach for the incoming reactive monomer 
is possible. 
The second cause is linked to the acceptance probability part as the tiny 
portion of space in which the reactive monomer is supposed to attach to 
maintain a limited additional intramolecular potential energy, will very often 
have a strong overlap with the wall highly repulsive region. 
We note that the wall potential energy climbs from $U_{\rm w}=0$ at the cutoff 
distance $x_{\rm c}/d=1.2$ from the wall up to $U_{\rm w}/k_{\rm B}T=5.8$ 
when the distance to the wall is decreased by $d/2$ at $x=0.7d$. 

The decrease of $U_i$ for $i \geq z$ can be estimated quantitatively starting from 
the micro-reversibility, Eq.~(\ref{eq:micro}), applied for arbitrary $k$, 
to $i=z+k-1$,
\begin{align}
U_{z+k-1}&=\frac{P_{z+k}}{P_{z+k-1}}W_{z+k}\nonumber \\ 
 &\approx \frac{\alpha_{z+k}}{\alpha_{z+k-1}} 
\hat{ \rho}_1^{\rm eff} W_{0} 
\approx \frac{\alpha_{z+k}}{\alpha_{z+k-1}}U_0
\label{eq:Upred}
\end{align}
where we have assumed that $W_{z+k}\approx W_0$ and approximated the size 
distribution according to Eq.~(\ref{eq:univ1}). 
Using the potential of mean force introduced in Eq.~(\ref{eq:potmf}), one has
\begin{equation}
U_{z+k-1}=\exp\left(- \beta (w_{z+k}-w_{z+k-1})\right)U_0
\label{eq:conf}
\end{equation}
which expresses that the rate of polymerization $U_{z+k-1}$ is the 
unperturbed rate $U_0$ times the Boltzmann factor implying the difference in 
confinement free energy for the same wall position, between filaments of size 
$z+k$ and $z+k-1$. Being specific, the use Eq.~(\ref{eq:Upred}) with 
$\alpha_z=1$ and other $\alpha_{z+k}$ 
values ($k \geq 1$) mentioned in Sec.~\ref{sec:simul_tech}, leads to 
$U_{15}=0.925 U_0$, $U_{16}=0.127 U_0$, $U_{17}=0.120 U_0$
and  $U_{18}=0.19 U_0$. 
These estimates are in quite good agreement with all data of experiment I 
in Fig.~\ref{fig:ee2b} which suggest a small systematic decrease of 
$U_{15}$ followed by a strong decrease for $U_{16}$ and $U_{17}$ followed 
by a slight but significant increase for $U_{18}$. 
The prediction of Eq.~(\ref{eq:Upred}) or of Eq.~(\ref{eq:conf}) that 
$U_i/U_0$ for $i \geq z$ is only dependent on the $\alpha$'s
is shown in inset in Fig.~\ref{fig:ee2b} and appears to remain valid in these 
slightly non-ideal conditions. 

The behavior of chemical rates in the other series of experiments remains 
similar to what was shown for Experiment I. 
However, one notes especially in the highest density series of 
experiments III, some systematic variations of $U_i$ and $W_i$ of the order 
of 20-30 percent within the observed size $i$ domain below the region of direct 
wall influence. 
This trend however coexists with the exponential distribution of the 
filament size probabilities Eq.~(\ref{eq:piexp}) and with the 
micro-reversibility requirement Eq.~(\ref{eq:micro}). 
The concept of unique rates $U_0$ and $W_0$ appears thus less appropriate in 
denser bundles and a more detailed analysis of these more complex inhomogeneous 
systems would be needed.

\section{Discussion and perspectives
\label{sec:disc}
}

Using the recently proposed simulation methodology to model the structure 
and dynamics of self-assembling stiff filaments at a mesoscopic level,
~\cite{Caby.12} we have investigated the dynamics of a bundle 
of grafted living filaments growing against a fixed wall.
In this Molecular Dynamics approach, each active end of the filament is 
subject to explicit local (de)polymerizations and monomer exchange with an 
embedding free monomer solution. 
Such a situation is reminiscent of cytoskeletal biofilament networks and more 
specifically of the actin filopodia where a set of parallel double-stranded 
actin filaments push on the cell membrane as a result of dominant 
end-filament polymerizations. 
However, being motivated by a generic molecular level understanding of the 
chemo-mechanical coupling, our strategy has been to exploit a minimum model 
of a confined solution of interacting spherical free monomers able, in 
presence of set of permanent grafted filament seeds, to dynamically 
self-assemble into an equivalent number of grafted wormlike chains with 
fluctuating contour length. 
For such model, crucial parameters like filament persistence length, chemical 
reaction rates, grafting density, diffusivity of free monomers, etc can be 
independently fixed and tuned.
This enables us to gain a general understanding of the physics 
behind the phenomena where chemical energy is transduced to develop 
mechanical forces and work, a basic phenomenon at the root of cell motility.

Specifically, for a large class of state points differing by the grafting 
density and by the total number of monomers in the confined volume, we have 
simulated a bundle of $32$ filaments grafted normally to a planar surface and 
hitting a fixed obstacle wall, parallel to the grafting wall, with gap width 
much smaller than the filament persistence length. 
For each state point, the simulations were invariably started from a set of 
$32$ grafted seeds and a state-point specific number of free monomers filling 
the available free space. 
As the simulation process is launched, the filaments start growing and the 
bundle spontaneously evolves towards an equilibrium situation for the 
bundle/free monomer solution system. 
Subsequent long microscopic trajectories were then exploited to probe the 
structure and some dynamical fluctuations of each investigated state point 
of this grafted bundle network.

As a first outcome of our work, the thermodynamic and statistical mechanics 
frameworks relevant to a bundle of living and interacting filaments in 
chemical equilibrium with a solution of free monomers is established in 
Sec.~\ref{sec:thermo}. 
Our approach, inspired by statistical mechanics concepts developed long time 
ago by T. L. Hill for reactive ideal and non-ideal mixtures,~\cite{b.Hill} is 
based on a reactive canonical ensemble and an associated free energy where 
both the number of filaments $N_f$ and the total number of monomers $N_t$ are 
fixed, in addition to the temperature $T$ and the geometrical dimensions of 
the confined volume, namely the gap width $L$ and transverse section area $A$. 
Section~\ref{sec:thermo1} introduces formally the correcting factors for 
intermolecular interactions, namely the activity coefficients $f_1$ (for free 
monomers) and $f_i$ (for grafted filaments of size $i$) and defines also a 
series of wall factors $\alpha_i(L)$ as correction terms to the ideal 
partition function of a single unconfined grafted filament of size $i$. 
To the free monomer chemical potential $\mu_1(N_t,N_f,T,A,L)$ in the bundle 
system can be associated a free monomer solution reservoir in equilibrium 
with the bundle solution, having thus a number density $\rho_1^{\infty}$ 
and $n$ free monomer activity coefficient $f_1^{\infty}$ related by 
$\mu_1=\ln{(f_1^{\infty}\rho_1^{\infty})}$. 
The osmotic force exerted by the bundle on the fixed wall, which is at the 
heart of the present work, must be strongly related to the distribution of 
filament sizes $P_i$. 
In Sec.~\ref{sec:thermo1}, the size probabilities are related by 
Eq.~(\ref{eq:equi_const3}) which summarizes how wall factors, activity 
coefficients of filaments and chemical potential $\mu_1$ of the free monomer 
reservoir in equilibrium with the bundle interfere to produce the equilibrium 
distribution of filament sizes. 
This framework is essential to discuss on equal footing theoretical 
predictions in the ideal bundle approach and all results obtained by our 
simulations in non-ideal conditions.

Precisely, a decisive advantage of our choice for a bundle network is that 
an exact treatment is possible for the corresponding ideal solution version 
of our confined bundle model. 
The latter follows by simply neglecting all intermolecular interactions 
while keeping unaltered from the full microscopic model, all intramolecular 
and wall interactions. 
The predictions for this ideal solution treatment (where all activity 
coefficients are set to unity), have already been detailed 
elsewhere.~\cite{RRMP.13} 
Given the importance of the ideal bundle case as a reference point, the 
salient properties of the ideal bundle case were summarized in 
Sec.~\ref{sec:thermo2}. 
The distribution of filament sizes is exponential 
$P_i \propto \exp{[ i \ln{(\rho_1 K_0)}]}$ as long as the sizes of the 
filaments are sufficiently small to avoid direct contact with the wall. 
In the size distribution expression, $K_0$ is the ideal solution 
(de)polymerization equilibrium constant of the active free end (in absence 
of confinement effects) and $\rho_1$ is the free monomer number density. 
Their product $\hat{\rho}_1=\rho_1 K_0$ provides the criterium for 
subcriticality $\hat{\rho}_1<1$ where depolymerization dominates or for 
supercriticality $\hat{\rho}_1>1$ where free ends of filaments are growing 
on average. 
In the supercritical case, the equilibrium size distribution is an 
increasing function of the size up to the point where the wall interactions 
start inhibiting the occurrence of longer filament sizes, their probability 
$P_i$ being rescaled by the corresponding wall factor $\alpha_i(L;l_{\rm p}))$ 
depending on the filament size, the gap width and the filament persistence 
length. 
The shape of this part of the distribution relative to filament sizes hitting 
the wall, results from a subtle balance between two opposite trends: the 
tendency for filaments to grow by polymerization and the price to pay for 
compressional free energy 
$w_i(L;l_{\rm p})= -k_{\rm B} T \ln{\alpha_i(L;l_{\rm p})}$ 
associated to the overall bending of filaments with contour length 
$L_{{\rm c}i}>L$. 
We note that the properties of these wall factors and their associated 
compression potentials of mean force have been recently studied for 
grafted wormlike chains hitting a hard wall.~\cite{Frey.06} 
In general, these wall factors (which need to be computed by single filament 
Monte-Carlo sampling~\cite{RRMP.13}) and the reduced free monomer density 
$\hat{\rho}_1$ provide the basic ingredients to establish at the same time 
the filament size distribution and the explicit expression of the compressive 
force $f_N^{\rm id}(L,\hat{\rho}_1)$ exerted by each independent filament on 
the fixed obstacle wall.

Returning now to the simulated non-ideal bundles, we first point out that for 
the different state points differing by the grafting density and by the total 
number of monomers, we have systematically estimated the equilibrium free 
monomer chemical potential $\mu_1$ by a particle insertion method adapted to 
the present reactive system which is derived in detail in the 
Supplementary Material.~\ref{sec:apA}
This allows to compare more easily bundle properties at different grafting 
density, in equilibrium with the same free monomer reservoir. 
For each state point, the structure of the bundle, the inhomogeneous free 
monomer density and the force exerted by the set of self-assembled stiff 
living filaments on the obstacle have been estimated. 
In addition, we have also recorded the effective rates of polymerization 
$U_i$ and depolymerization $W_i$ for both short filaments (not hitting the 
wall) and for long filaments which are directly in contact with the wall. 
We have analyzed how these rates are influenced by the interactions between 
filaments and free monomers and by the obstacle confinement. 
This large set of results, regrouped in Tables~\ref{tab.2},~\ref{tab.3},
~\ref{tab.4} and~\ref{tab.5} are produced at many state points on the basis 
of a unique microscopic model for which we know exactly the properties 
within the ideal solution approximation.
This allows us to highlight the various influences of interactions or of 
confinement on the living filament bundle properties and this analysis is a 
major original outcome of our work.

The main result of our simulations on non ideal bundles is that the same 
functional form applies in ideal and non-ideal situations, when studying the 
dependence of structural properties on the free monomer density. 
The explicitly known $\hat{\rho}_1$ dependence of the wall force per filament 
in the ideal case at a given gap width $L$, namely 
$f_N^{\rm id}=\phi(\hat{\rho}_1;L,T)$, turns out to be valid to describe the 
force per filament for a bundle with strong inter-filament and/or strong 
filament-free monomers interactions, provided $\hat{\rho}_1$ is replaced by an 
appropriate renormalized reduced free monomer density 
$\hat{\rho}_1^{\rm eff}$. 
Pragmatically estimated from the exponential size distribution of filaments 
which are not interacting with the wall, its thermodynamic interpretation 
could be traced to the combination 
$\hat{\rho}_1^{\rm eff}=K_0 f_1^{\infty}\rho_1^{\infty}f_{i-1}/f_i$ 
where the ratio $f_{i-1}/f_i$ is the effectively $i$-independent quantity 
within the free (non hitting) filaments regime. 
With our choice of purely repulsive intermolecular interactions between 
filaments, the average wall force per filament appears to decrease as the 
grafting density increases for a series of bundles exposed to the same free 
monomer reservoir chemical potential. 
This property is coherent with the opposite trend predicted theoretically for 
a bundle of attractive filaments.~\cite{Krawczyk.11} 
In summary, we observed in our simulated bundles in supercritical conditions, 
that the repulsive interactions weaken the effective polymerization tendency 
by slowing down the static exponential growth of the short filaments 
distribution, an effect directly coupled to the similar effective weakening of 
the wall osmotic pressure exerted by the filaments.

Coming back to general aspects valid for both ideal and non ideal bundles, 
it must be stressed that the wall forces per filament $f_N$ discussed in this 
work concern a unique and (ensemble) fixed value of $L$. 
Ideal bundle theory~\cite{RRMP.13} indicated that the force 
$f_N^{\rm id}(L;\hat{\rho}_1,T)$ for a single filament varies quite strongly 
with $L$ over a single monomer size interval $L-d/2,L+d/2$, basically as a 
result of the strong variations with $L$ of the populations of the different 
filament chemical species hitting the wall and as a result of the variation of 
the wall factors $\alpha_i$ with $L$. 
These variations of $f_N$ tend to be replicated periodically in space with a 
period $d$ at larger $L$ values, but with some systematic damping effect 
related to the progressive increase of the ratio $L/l_{\rm p}$. 
However, if the force is averaged over a $L$ window of size $d$, a result close 
to Eq.~(\ref{eq:force}) is recovered~\cite{RRMP.13} suggesting that the 
force $F$ mentioned in ratchet theories for the stalling force needed to 
stop polymerizing filaments~\cite{b.Howard} must be interpreted as an average 
force whose work $Fd$ over a distance $d$ has to be identified with the 
integral over a $L$ interval of size $d$ of the fixed $L$ equilibrium force 
discussed we have been measuring in our simulations. 
Similarly, the $L$ dependent force cannot be related directly to the force 
measured in an optical trap experiment~\cite{Dogterom.07} where a mobile wall 
(bead) is subject to both the bundle pressure and the restoring trapping force. 
The unavoidable Brownian motion of the confining mobile wall (bead) in the trap 
(of the order of $\sqrt{k_{\rm B}T/k_{\rm trap}}$ lead to fluctuations of the 
effective size of the tested bundle which are larger than $d$ (in the 
experiment on actin, $d=2.7$ nm and $\sqrt{k_{\rm B}T/k_{\rm trap}} \cong23$ nm),
 hence the quantity measured should again be an average of the filament wall 
force over a range larger than $d$. 
In principle, according to our results on non-ideal bundles, a plausible 
estimate of such an average wall force per filament in a non-ideal case should 
approach $(k_{\rm B}T/d )\ln{(\hat{\rho}_1^{\rm eff})}$.

Finally, our simulations give access to kinetics aspects for chemical reactions 
involving the grafted filaments. 
Given our algorithm where attempted reaction steps are subsequently either 
refused or accepted according to the associated microscopic energetic changes, 
the polymerization and depolymerization rates are affected by both the 
intermolecular interactions and by the confinement potential. 
We found in the low grafting surface density case that the rates are 
$i$-independent in the short (non hitting) filament regime, but are strongly 
dependent on the state point with the property 
$U_0/W_0=\hat{\rho}_1^{\rm eff}$ (where $U_0$ and $W_0$ denote these 
polymerizing and depolymerizing homogeneous rates respectively), which is a 
direct consequence of microreversibility and the insensitivity to $i$ of the 
rates, given the exponential filament size distribution profile discussed 
earlier.
The wall effect on these rates for filament sizes hitting the wall was 
analyzed and found to be rather weak for depolymerization but on the 
contrary spectacular for the polymerization rates. 
A first order approximation for these rates for hitting filament sizes was 
suggested as $W_i=W_0$ and $U_i=(\alpha_{i+1}/\alpha_i)U_0$, which 
again take into account both the state point influence (through the bulk 
rates $U_0$ and $W_0$) and the wall influence (through the wall factors).

In conclusion, the ensemble of structural and dynamic results we have 
obtained for a system of grafted and interacting living filaments at 
equilibrium can be rationalized in terms of a reference state, namely an 
ideal confined bundle of living filaments, to which non-ideality corrections 
can be identified. 
Beyond this important point, the methodology followed to get these 
results suggests that we have at disposal a robust machinery able to 
drive our living system spontaneously, thanks to a combination of free 
monomer diffusive steps and (de)polymerizing chemical steps, to the 
equilibrium state of a living filament network, under the constraints of a 
well defined statistical mechanics ensemble. 
It suggests that the methodology could be extended to more complex 
filament networks (e.g. branched networks, flexible membranes obstacles, 
distinction between hydrolyzed and non hydrolyzed actin complexes etc) at 
the price of an increase in the number of chemical species for the building 
blocks and at the price of the consideration of a larger set of competing 
chemical reactions. 
Such complex networks are already studied by 
biophysicists,~\cite{Ca.01,mogilner.96,mogilner.05} 
generally in non-equilibrium conditions, on the basis of ad hoc stochastic 
rules regarding various probabilities for filament branching, filament capping 
or filament elongation by polymerization, and on the basis of hypotheses on how 
a wall force affects these probabilities and how the load force exerted on the 
wall must be distributed among all hitting filaments. 
These stochastic network simulations usually deal with wider non-equilibrium 
dynamical properties like the force-velocity relationship for a network 
pressing against a mobile wall. 
Our methodology is ripe not only for extensions towards more complex networks 
(at equilibrium) but can also be applied to dynamical network properties by 
releasing the fixed wall constraint while exerting an external load instead. 
Work along these lines are in progress. 
It should help interpreting the chemo-mechanical coupling for a simple network 
of parallel living filaments pushing on a mobile wall, a problem addressed in 
recent theoretical works, either directly for the bundle 
case~\cite{Joanny.11} or for more complex actin networks.~\cite{Ca.01}

\begin{acknowledgments}
The authors wish to thank M. Baus, G. Ciccotti, J.-F. Joanny, P. B. S. Kumar, 
D. Lacoste and C. Pierleoni for useful discussions about the present work. 
They warmly thank G. Destr\'{e}e for invaluable technical help. 
S. Ramachandran acknowledges financial help from the BRIC 
(Bureau des Relations Internationales et de Coop\'eration) of the 
Universit\'e Libre de Bruxelles. 
J.-P. R. thanks P. B. S. Kumar for hosting him at IIT Madras where this work 
was completed. 
\end{acknowledgments}

\appendix

\renewcommand{\theequation}{A. \arabic{equation}}\setcounter{equation}{0}

\section{Derivation of the chemical potential $\mu_1$ of free monomers 
in chemical equilibrium with the anchored bundle within the reacting 
canonical ensemble
\label{sec:apA}
}

This appendix establishes the final expression given by Eq.~(41)
which allows us to compute operationally, via a Widom-like method, the chemical 
potential of the free monomers within our reacting mixture at chemical 
equilibrium. 
From the formal definition in Eq.~(10), we can write
$\mu_1=F(N_t+1,N_f,A,L,T)-F(N_t,N_f,A,L,T)$,
and thus, using the short symbol $Q(N_t,X)$ for the reactive canonical 
ensemble partition function where $X$ denotes the remaining set of 
independent variables $X=(N_f,A,L,T)$, one has
\begin{align}
\mu_1(N_t,X)
&=-k_{\rm B}T 
\ln \frac{Q(N_t+1,X)}{Q(N_t,X)}, \label{eq:mui}\\
Q(N_t,X)&= {\sum_{N_1,\{N_i\}}^{N_t}} 
\frac{1}{h^{3N_t} N_1!\ldots N_i!\ldots} 
\int {\rm d}\Gamma^{N_t} \exp{[-\beta H(N_1,\{N_i\})]},\label{eq:QNT}\\
Q(N_t+1,X)&= {\sum_{N_1^{'},\{N_i^{'}\}}^{N_t+1}}^{'} 
\frac{1}{h^{3(N_t+1)}N_1^{'}!\ldots N_i^{'}!\ldots} 
\int {\rm d}\Gamma^{N_t+1} 
\exp{[-\beta H(N_1^{'},\{N_i^{'}\})]},
\label{eq:QNT1}
\end{align}
where the sum in Eq.~(\ref{eq:QNT}) and the sum in Eq.~(\ref{eq:QNT1}) imply 
in each case the sum over all distinct sets of integer values (starting from 0) 
such that the total number of filaments is $N_f$ while the total number of 
monomers is $N_t$ in the first case and $N_t+1$ in the second case. 
Note that the possible values for the number of species of the same kind are 
represented by unprimed quantities for $N_t$ ($N_1, N_3,\ldots$) and by 
primed quantities for $N_t+1$ ($N_1', N_3',\ldots$).

To illustrate next developments, the table below provides the complete 
set of distinct arrangements of populations satisfying $N_f=2$ and either 
$N_t=9$, $N_t=10$ or $N_t=11$. 
This table shows that if one knows all distinct sets ($N_1,N_3,\ldots$) 
for one pair of values of $N_f$ and $N_t$, the whole set of possibilities 
$N_1', N_3',\ldots$ corresponding to the pair of values $N_f$ and $N_t+1$, 
can be divided into two classes
\begin{itemize}
\item those for which $N_1' \geq 1$ which are in one-to-one correspondence 
with the same set of filament populations in the previous column for 
$N_t$ arrangements, with $N_3'=N_3$,\ldots, $N_i'=N_i \ldots$,  
while $N_1'=N_1+1$.
\item those for which $N_1'=0$ which correspond to all distinct topological 
ways to distribute $N_t+1$ monomers into $N_f$ filaments without any 
free monomer left.
\end{itemize}

On the basis of this observation, one can rewrite Eq.~(\ref{eq:mui}) for a 
set of $N_t+1$ monomers (in terms of $N_1',N_3',\ldots$ sets) in separating 
the two types of contributions distinguished above, the first term being 
rewritten automatically as a sum over all distinct population sets of the 
$N_t$ case (in terms of the $N_1,N_3,\ldots$ sets and the second being limited 
to all sets with $N_1'=0$. 
This gives
\begin{align}
Q(N_t+1,X)&= Q_1+Q_2, \label{eq:QQ}
\end{align}
\begin{align}
&Q_1 = {\sum_{N_1,\{N_i\}}^{N_t}} 
\frac{1}{h^{3(N_t+1)} (N_1+1)! \ldots N_i!\ldots} 
\int {\rm d}\Gamma^{N_t+1} \exp{[-\beta H(N_1+1,\{N_i\})]}, 
\label{eq:QQ1}
\end{align}
\begin{align}
Q_2 &= {\sum_{\{N_i^{'}\}}^{N_t+1}} 
\frac{1}{h^{3(N_t+1)} N_3^{'}\ldots N_i^{'}!\ldots} 
\int {\rm d}\Gamma^{N_t+1} 
\exp{[-\beta H(N_1^{'}=0,\{N_i^{'}\})]}
\label{eq:QQ2}
\end{align}
where the sum in $Q_2$ implies a number of $N_f$ filaments of size 
between $i=3$ and $i=z^{*}$ and a total number of monomers $N_t+1$ 
satisfying $N_t+1=\sum_i i N_i'$.
We now need to exploit the partition indicated in Eq.~(\ref{eq:QQ}) to evaluate 
the ratio of partition functions $Q_1/Q$ and $Q_2/Q$ (see Eq.~(\ref{eq:mui})) 
and we now treat both cases successively.

Looking at $Q_1$ as defined by Eq.~( \ref{eq:QQ1}), we isolate in each term 
of the sum one free monomer (considered as an extra monomer which will be 
specified by six coordinates written as $\Gamma_{\rm ex}$) in the integrals 
and we also isolate the corresponding spatial coordinate in the 
Hamiltonian where $\Delta H_{\rm ex}$ contains all kinetic and potential 
terms which need to be added when this extra particle is incorporated to the 
system,  
\begin{align}
Q_1=& {\sum_{N_1,\{N_i\}}^{N_t}} 
\int {\rm d}\Gamma^{N_t} \frac{1}{h^3 (N_1+1)} 
   \int {\rm d}\Gamma_{\rm ex} 
\exp{[-\beta \Delta H_{\rm ex}]} 
\frac{1}{h^{3N_t} N_1!\ldots N_i!\ldots} 
\exp{[-\beta H(N_1,\{N_i\})]}.
\end{align}
Inspection of the ratio $Q_1/Q(N_t,X)$ indicates that it corresponds to an 
equilibrium average over the $(N_t,X)$ ensemble, namely
\begin{align}
 \frac{Q_1}{Q(N_t,X)}&= 
\left\langle 
\frac{1}{h^3 (N_1+1)} 
\int {\rm d}\Gamma_{\rm ex} 
\exp{[-\beta \Delta H_{\rm ex}]}
\right\rangle_{N_t,X}, \\
&= \frac{V (2 \pi M k_{\rm B}T)^{3/2}}{h^3} 
\left\langle 
\frac{1}{(N_1+1)} 
\int_0^1 {\rm d}s_{\rm ex}^3 
\exp{[-\beta \Delta U_{\rm ex}]}\right\rangle_{N_t,X},\\
&= \frac{V}{\Lambda^3} 
\left\langle \frac{1}{(N_1+1)} 
\int_0^1 {\rm d}s_{\rm ex}^3 
\exp{[-\beta \Delta U_{\rm ex}]}\right\rangle_{N_t,X},
\label{eq:az}
\end{align}
where the integrations over the extra particle degrees have been either 
performed explicitly for the momenta or written in terms of reduced 
coordinates ${\bf s}_{\rm ex}$ (using a reduction by the box dimensions) 
for their coordinates. 
$\Delta U_{\rm ex}$ is the potential energy increase when the extra particle 
is added to the set of $N_t$ monomers of the system associated to the ensemble 
on which average is taken. 
Finally, $M$ is the monomer mass and $\Lambda$ is the de Broglie wave length.

$Q_2$ in Eq.~(\ref{eq:QQ2}) has the structure of a partition function 
relative to a subset of microscopic configurations of the original ensemble 
that could be written as $Q^0(N_t+1,X)$.
Here the superscript indicates that only the subset of microscopic 
configurations with $N_1=0$ are included. 
Using this new notation, we rewrite the ratio $Q_2/Q$ as 
\begin{align}
Q_2/Q=& \frac{Q^0(N_t+1,X)}{Q(N_t,X)} 
   = \frac{Q^0(N_t+1,X)}{Q^0(N_t,X)}\frac{Q^0(N_t,X)}{Q(N_t,X)} \\
     =& \frac{Q^0(N_t+1,X)}{Q^0(N_t,X)} P(N_1=0;N_t,X).
\end{align}
In the thermodynamic limit, the probability $P(N_1=0;N_t,X)$ for our system 
in chemical equilibrium to be in any state where $N_1=0$ is vanishingly small.
 We checked that in our bundle simulations with a few hundreds monomers, 
the fluctuation $\delta N_1$ around the mean $\langle N_1\rangle $ remains 
typically of the order of $10-15$ percent of this mean, indicating that 
finite size effects on $P(N_1=0;N_t,X)$ remain negligible. 
We note that this very small probability is multiplied by a term which has 
the same structure as the $Q_1/Q$ term and which could be expressed in terms 
of the insertion probability of a filament of size $i+1$ in replacement of a 
filament of size $i$ in the sub-ensemble of cases where $N_1=0$. 
Concluding, if we neglect the $Q_2$ contribution, we get the final expression
\begin{align}
\beta \mu_1(N_t,X)&=
- \ln \frac{Q_1+Q_2}{Q(N_t,X)}\cong - \ln \frac{Q_1}{Q(N_t,X)},
\end{align}
so that using Eq.~(\ref{eq:az}), 
Eqs.~(40) and~(41)
are recovered.

\begin{table}
\caption{List of distinct population possibilities for a reacting system 
with a total of $N_t$ monomers and a total of $N_f$ filaments arranged in 
$N_1$ free monomers, $N_3$ filaments of length $3, \ldots, N_i$ filaments of 
length $i,\ldots$ for three successive values of $N_t$ and same $N_f$. 
$N_1$ values are indicated explicitly, while only non zero populations of 
filaments of size $i$ are indicated. 
All distinct possibilities are grouped for one specific value of $N_1$ in 
order to appreciate the links between sets relative to consecutive values of 
$N_t$.}
\begin{center}
\begin{tabular}{ccc}
\hline
\hline
&&\\
$N_t=9$  $N_f=2$ 
& $N_t=10$  $N_f=2$ 
& $N_t=11$  $N_f=2$ 
\\
&&\\
\hline
&&\\
&
&
$
N_1=0
  \begin{cases}
  N_5=N_6=1 \\
  N_4=N_7=1 \\
  N_3=N_8=1 
  \end{cases}
$
\\
&&\\
\hline
&&\\
&
$
N_1=0
  \begin{cases}
  N_5=2 \\
  N_4=N_6=1 \\
  N_3=N_7=1 
  \end{cases}
$
&
$
N_1=1
  \begin{cases}
  N_5=2 \\
  N_4=N_6=1 \\
  N_3=N_7=1 
  \end{cases}
$
\\
&&\\
\hline
&&\\
$
N_1=0
  \begin{cases}
  N_3=N_6=1 \\
  N_4=N_5=1 
  \end{cases}
$
&
$
N_1=1
  \begin{cases}
  N_3=N_6=1 \\
  N_4=N_5=1 
  \end{cases}
$
&
$
N_1=2
  \begin{cases}
  N_3=N_6=1 \\
  N_4=N_5=1 
  \end{cases}
$
\\
&&\\
\hline
&&\\
$
N_1=1
  \begin{cases}
  N_4=2 \\
  N_3=N_5=1 
  \end{cases}
$
&
$
N_1=2
  \begin{cases}
  N_4=2 \\
  N_3=N_5=1 
  \end{cases}
$
&
$
N_1=3
  \begin{cases}
  N_4=2 \\
  N_3=N_5=1 
  \end{cases}
$
\\
&&\\
\hline
&&\\
$
N_1=2 \; N_3=N_4=1 
$
&
$
N_1=3 \; N_3=N_4=1
$
&
$
N_1=4 \; N_3=N_4=1 
$
\\
&&\\
\hline
&&\\
$
N_1=3 \; N_3=2
$
&
$
N_1=4 \; N_3=2
$
&
$
N_1=5 \; N_3=2
$
\\
&&\\
\hline
\hline
\end{tabular}
\end{center}
\end{table}

\newpage
\nocite{*}
%

\end{document}